\documentclass[twocolumn]{aastex631}

\usepackage{hyperref}
\usepackage{multirow}
\usepackage{booktabs}
\usepackage{array}
\usepackage{rotating}
\newcolumntype{C}{@{\extracolsep{3cm}}c@{\extracolsep{0pt}}}%
\newcommand{\nefive}{[\ion{Ne}{5}]$\,\lambda$14.32\,$\mu$m}
\newcommand{\ofour}{[\ion{O}{4}]$\,\lambda$25.91\,$\mu$m}
\newcommand{\nethree}{[\ion{Ne}{3}]$\,\lambda$15.56\,$\mu$m}
\newcommand{\netwo}{[\ion{Ne}{2}]$\,\lambda$12.81\,$\mu$m}
\newcommand{\fetwo}{[\ion{Fe}{2}]$\,\lambda$5.34\,$\mu$m}

\begin{document}

\title{ReveaLLAGN 0: First Look at JWST MIRI data of Sombrero and NGC~1052}

\author[0000-0002-7743-9906]{Kameron Goold}
\affiliation{Department of Physics \& Astronomy, University of Utah, James Fletcher Building, 115 1400 E, Salt Lake City, UT 84112, USA}

\author[0000-0003-0248-5470]{Anil Seth}
\affiliation{Department of Physics \& Astronomy, University of Utah, James Fletcher Building, 115 1400 E, Salt Lake City, UT 84112, USA}

\author[0000-0001-8440-3613]{Mallory Molina}
\affiliation{Department of Physics \& Astronomy, University of Utah, James Fletcher Building, 115 1400 E, Salt Lake City, UT 84112, USA}
\affiliation{Department of Physics \& Astronomy, Vanderbilt University, Nashville, TN 37235, USA}

\author[0009-0004-9457-2495]{David Ohlson}
\affiliation{Department of Physics \& Astronomy, University of Utah, James Fletcher Building, 115 1400 E, Salt Lake City, UT 84112, USA}

\author[0000-0001-8557-2822]{Jessie C. Runnoe}
\affiliation{Department of Physics \& Astronomy, Vanderbilt University, Nashville, TN 37235, USA}


\author[0000-0002-5666-7782]{Torsten B\"oker}
\affiliation{European Space Agency, c/o STScI, 3700 San Martin Drive, Baltimore, MD 21218, USA}

\author[0000-0003-4932-9379]{Timothy A. Davis}
\affiliation{Cardiff Hub for Astrophysics Research \& Technology, School of Physics \& Astronomy, Cardiff University, Queens Buildings, Cardiff, CF24 3AA, UK}

\author[0000-0003-0234-3376]{Antoine Dumont}
\affiliation{Max-Planck-Institut f{\"u}r Astronomie, K{\"o}nigstuhl 17, D-69117, Heidelberg, Germany}

\author[0000-0002-3719-940X]{Michael Eracleous}
\affiliation{Department of Astronomy \& Astrophysics and Institute for Gravitation and the Cosmos, The
Pennsylvania State University, 525 Davey Lab, University Park, PA 16802, USA}

\author[0000-0001-9490-899X]{Juan Antonio Fern\'{a}ndez-Ontiveros}
\affiliation{Istituto di Astrofisica e Planetologia Spaziali (INAF–IAPS), Via Fosso del Cavaliere 100, I–00133 Roma, Italy}
\affiliation{Centro de Estudios de F\'{i}sica del Cosmos de Arag\'{o}n (CEFCA), Plaza San Juan 1, E–44001, Teruel, Spain}

\author[0000-0001-5802-6041]{Elena Gallo}
\affiliation{Department of Astronomy, University of Michigan, 1085 S. University Ave., Ann Arbor, MI 48109, USA}

\author[0000-0003-4700-663X]{Andy D. Goulding}
\affiliation{Department of Astrophysical Sciences, Princeton University, Princeton, NJ 08544, USA}

\author[0000-0002-5612-3427]{Jenny E. Greene}
\affiliation{Department of Astrophysical Sciences, Princeton University, Princeton, NJ 08544, USA}

\author[0000-0001-6947-5846]{Luis C. Ho}
\affiliation{Kavil Institute for Astronomy and Astrophysics, Peking University, Beijing 100871, China}
\affiliation{Department of Astronomy, School of Physics, Peking University, Beijing 100871, China}

\author[0000-0001-9564-0876]{Sera B. Markoff}
\affiliation{Anton Pannekoek Institute for Astronomy, University of Amsterdam, Science Park 904, 1098 XH Amsterdam, The Netherlands}

\author[0000-0002-6922-2598]{Nadine Neumayer}
\affiliation{Max-Planck-Institut f{\"u}r Astronomie, K{\"o}nigstuhl 17, D-69117, Heidelberg, Germany}

\author[0000-0002-7092-0326]{Richard M. Plotkin}
\affiliation{Department of Physics, University of Nevada, Reno, NV 89557, USA}
\affiliation{Nevada Center for Astrophysics, University of Nevada, Las Vegas, NV 89154, USA}

\author[0000-0002-3585-2639]{Almudena Prieto}
\affiliation{Universidad de La Laguna (ULL), Dpto. Astrof\'{i}sica, Avd. Astrof\'{i}sico Fco. S\'{a}nchez s/n, 38206 La Laguna, Tenerife, Spain}
\affiliation{Instituto de Astrof\'{i}sica de Canarias (IAC), C/V\'{i}a L\'{a}ctea s/n, 38205 La Laguna, Tenerife, Spain}
\affiliation{Universit{\"a}ts-Sternwarte, Fakult{\"a}t f{\"u}r Physik, Ludwig-Maximilians-Universit{\"a}t M{\"u}nchen, 81679 M{\"u}nchen, Germany}

\author[0000-0003-2277-2354]{Shobita Satyapal}
\affiliation{George Mason University, Department of Physics and Astronomy, MS3F3, 4400 University Drive, Fairfax, VA 22030, USA}

\author[0000-0003-4546-7731]{Glenn van de Ven}
\affiliation{Department of Astrophysics, University of Vienna, T\"urkenschanzstra{\ss}e 17, 1180 Vienna, Austria}

\author[0000-0002-1881-5908]{Jonelle L. Walsh}
\affiliation{George P. and Cynthia W. Mitchell Institute for Fundamental Physics and Astronomy, Department of Physics \& Astronomy, Texas A\&M University, 4242 TAMU, College Station, TX 77843, USA}

\author[0000-0003-3564-6437]{Feng Yuan}
\affiliation{Shanghai Astronomical Observatory, Chinese Academy of Sciences, Shanghai 200030, People’s Republic of China}


\author[0000-0002-0160-7221]{Anja Feldmeier-Krause}
\affiliation{Max-Planck-Institut f{\"u}r Astronomie, K{\"o}nigstuhl 17, D-69117, Heidelberg, Germany}

\author[0000-0002-1146-0198]{Kayhan G{\"u}ltekin}
\affiliation{Department of Astronomy, University of Michigan, 1085 S. University Ave., Ann Arbor, MI 48109, USA}

\author[0000-0002-6353-1111]{Sebastian H{\"o}nig}
\affiliation{Department of Physics \& Astronomy, University of Southampton, Hampshire SO17 1BJ Southampton, UK}

\author[0000-0002-5537-8110]{Allison Kirkpatrick}
\affiliation{Department of Physics and Astronomy, University of Kansas, Lawrence, KS 66045, USA}

\author[0000-0002-4034-0080]{Nora L{\"u}tzgendorf}
\affiliation{European Space Agency, c/o STScI, 3700 San Martin Drive, Baltimore, MD 21218, USA}

\author[0000-0001-7158-614X]{Amy E. Reines}
\affiliation{eXtreme Gravity Institute, Department of Physics, Montana State University, Bozeman, MT 59717, USA}

\author[0000-0002-1468-9668]{Jay Strader}
\affiliation{Department of Physics and Astronomy, Michigan State University, East Lansing, MI 48824, USA}

\author[0000-0002-1410-0470]{Jonathan R. Trump}
\affiliation{Department of Physics, 196 Auditorium Road, Unit 3046, University of Connecticut, Storrs, CT 06269, USA}

\author[0000-0001-6215-0950]{Karina T. Voggel}
\affiliation{Universite de Strasbourg, CNRS, Observatoire astronomique de Strasbourg, UMR 7550, 67000 Strasbourg, France}



\begin{abstract}

We present the first results from the Revealing Low-Luminosity Active Galactic Nuclei (ReveaLLAGN) survey, a JWST survey of seven nearby LLAGN.  We focus on two observations with the Mid-Infrared Instrument's (MIRI) Medium Resolution Spectrograph (MRS) of the nuclei of NGC~1052 and Sombrero (NGC~4594 / M104).  We also compare these data to public JWST data of a higher-luminosity AGN, NGC~7319 and NGC~7469.   JWST clearly separates the AGN spectrum from the galaxy light even in Sombrero, the faintest target in our survey; the AGN components have very red spectra.   We find that the emission-line widths in both NGC~1052 and Sombrero increase with increasing ionization potential, with FWHM$>$1000~km$\,$s$^{-1}$ for lines with ionization potential $\gtrsim$ 50~eV.  These lines are also significantly blue-shifted in both LLAGN.  The high ionization potential lines in NGC~7319 show neither broad widths or significant blue shifts.  Many of the lower ionization potential emission lines in Sombrero show significant blue wings extending $>$1000~km$\,$s$^{-1}$.  These features and the emission-line maps in both galaxies are consistent with outflows along the jet direction.  Sombrero has the lowest luminosity high-ionization potential lines ([\ion{Ne}{5}] and [\ion{O}{4}]) ever measured in the mid-IR, but the relative strengths of these lines are consistent with higher luminosity AGN.  On the other hand, the [\ion{Ne}{5}] emission is much weaker relative to the [\ion{Ne}{3}] and [\ion{Ne}{2}] lines of higher-luminosity AGN.  These initial results show the great promise that JWST holds for identifying and studying the physical nature of LLAGN.  

\end{abstract}



\section{Introduction} \label{sec:intro}

As material falls onto a black hole, that material heats up and emits light creating an active galaxy nucleus (AGN).  While the most rapidly accreting objects are seen to the edges of our Universe as luminous quasars, the vast majority of central supermassive black holes in nearby galaxies are accreting at less than 1\% of their Eddington Limit \citep[$L_{\rm bol}/L_{\rm Edd} < 0.01$; see][]{Ho2009}. These low-luminosity AGN (LLAGN) are theorized to have significantly different inner structures from the accretion disks found in more luminous AGN. 

At these low accretion rates, the inner part of the optically thick accretion disk transitions to a hot, optically thin, radiatively inefficient accretion flow \citep[RIAF;][]{Narayan1995, Yuan2014,  Porth2019}. This change in the central regions of LLAGN will result in a different ionizing spectrum with fewer far-ultraviolet photons.  

Observationally, this is confirmed by the lack of the ``big blue bump" in LLAGN spectral energy distributions (SEDs; \citealt{ho1999}).  This change in ionizing flux is also expected to be reflected in the optical emission line strengths. Enhanced low ionization emission lines are a key characteristic of low ionization nuclear emission regions (LINERs) which were first identified by \cite{Heckman1980} based solely on optical oxygen lines. LINERs are notably diverse, including sources both with and without clear evidence of an AGN. Multiple radio and X-ray surveys have consistently revealed that most LINERs are powered by LLAGNs \citep{Nagar2002, Nagar2005, Filho2006, Dudik2005, Flohic2006, gonzalez2006, gonzalez2009, Ho2008, hernandez2013, hernandez2014}. However, LLAGNs are not coincident with LINERs exclusively, many weakly accreting Seyferts are also considered LLAGNs \citep{Kewley2006, Ho2009}. Optical classification not withstanding, LLAGNs share additional observational signatures. In particular, the dusty torus and broad line region components may disappear \citep[e.g.][]{Plotkin2012, Elitzur2014}; and as the Eddington Ratio decreases, LLAGN tend to have stronger jet emission \citep{Ho2008} and become increasingly radio-loud \citep{Ho2002, Terashima2003, Greene2006, panessa2007, sikora2007, Trump2011}. The kinetic energy injected into LLAGN host galaxies by jets may play a significant role in keeping massive early-type galaxies quiescent \citep{Croton2006, Weinberger2017}. Despite these observational signatures the inner structure of LLAGNs are still not yet well understood and it becomes increasingly difficult to separate out the low luminosity nuclear emission of weakly accreting AGN from the surrounding light and obscuring dust of the host galaxy.

Infrared (IR) wavelengths are particularly valuable for studying AGN \citep{Sajina2022}, as the dust that hides many AGN at optical and UV wavelengths strongly emits in the IR.  In fact, the energy output for many AGN is highest at X-ray and mid-IR wavelengths \citep{Prieto2010}.  Furthermore, the emission from AGN at 12~$\mu$m has been found to be tightly correlated with the 2-10 keV X-ray emission, with similar luminosities in both bands \citep{Asmus2015}. In addition to the continuum emission from dust or jet emission \citep[e.g.][]{Prieto2016,Fern2023}, strong emission lines are seen at infrared wavelengths, including high ionization potential (IP) ``coronal'' emission lines that track the ionizing spectrum of the AGN \citep[e.g.][]{Satyapal2008,Goulding2009}.  

JWST, operating primarily in the IR, is equipped with advanced instruments and brings new opportunities in the study of AGN. The brightness of AGN in the IR beyond 2 microns combined with JWST’s unprecedented sensitivity at these wavelengths makes it the most sensitive instrument ever for detecting AGN. For example, the depth reached in just 10 ks of Mid-Infrared Instrument (MIRI) imaging at 12 microns roughly matches that of 2~Ms from Chandra Deep Field North \citep[][assuming the \citet{Asmus2015} relation between the mid-IR and X-ray emission]{Xue2016}. The remarkable spatial resolution afforded by JWST's 6.5-meter diameter mirror allows us to isolate the LLAGN emission from that of the host galaxy in nearby objects.  Finally, JWST's spectral resolution enables studies of line emission profiles that were not possible with previous missions.  

The Revealing LLAGN (ReveaLLAGN) project, utilizing integral field spectroscopic (IFS) observations from JWST, aims to achieve two primary goals. The first is to provide templates of LLAGN spectra, which can be used to identify the abundant faint AGN hidden in future JWST data of local and high-redshift galaxies.  This includes environments where their presence is currently uncertain, e.g. in dwarf galaxies. Second, through the analysis of the continuum and coronal-line emissions, the project aims to offer valuable constraints for understanding the internal structure of LLAGN. The study focuses on seven nearby, well-known LLAGN covering a wide range of both black hole mass (10$^{5.5\textrm{--}9.8}$~M$\odot$) and Eddington ratio (log($L_{bol}/L_{edd}$) ranging from -6.2 to -2.7).

In this paper, we report the first results from the ReveaLLAGN project based on the MIRI medium-resolution spectrometer (MRS) data from our first two targets, Sombrero (also known as M104 and NGC~4594) and NGC~1052. The overall properties of these galaxies are listed in Table~\ref{table:galprop}.  These two galaxies have the highest (NGC~1052) and lowest (Sombrero) 12~$\mu$m fluxes \citep{asmus2014} of all the galaxies in the full ReveaLLAGN sample (Seth et al., {\em in prep}), and thus represent the full range of signal-to-noise ratios (S/N) expected for the survey. NGC~1052 and Sombrero are classified as LINERs based on their optical emission lines \citep{Heckman1980, Ho1997}\footnote{We note that the line ratios of NGC~1052 depends on radius, and are Seyfert-like at smaller radii \citep{Molina2018}} and exhibit extensive multiwavelength emission from their LLAGN. Hard X-ray observations reveal point sources in the center of both galaxies (\emph{NGC1052}: \citealt{Guainazzi1999, Kadler2004}; \emph{Sombrero}: \citealt{Fabbiano1997, Pellegrini2002, Pellegrini2003}), accompanied by UV variability \citep{maoz2005}. In the radio domain, NGC~1052 hosts jets at parsec scales with a position angle of $\sim$70 degrees \citep{claussen1998, Kadler2004}, while at kiloparsec scales the PA of the radio jets are seen at $\sim$100 degrees \cite{Wrobel1984, Kadler2004a}. The Sombrero Galaxy also contains compact jets, observed at sub-parsec scales with a PA of -25 degrees \citep{Hada2013}. Additionally, both AGNs' SEDs show a lack of emission in the UV relative to higher luminosity AGN \citep[][]{Fern2023}, consistent with other LLAGN \citep{Ho2008}.  We review previous observations of both galaxies' AGN in more depth in Section~\ref{sec:outflow_discussion}.

We contrast these two LLAGN observations with previous Spitzer data of higher luminosity AGN.  We also include JWST MIRI/MRS observations of NGC~7319 \citep[part of the JWST Early Release Observations;][]{Ponto2022} and NGC~7469 \citep{Armus2023}, two Seyfert galaxies with higher luminosity and Eddington ratios than our targets.  
The 2-10 keV X-ray luminosities of NGC~7319 and NGC~7469 are  $10^{43.1}$ erg s$^{-1}$ and $10^{43.2}$ erg s$^{-1}$ \citep{Ricci2017}. Their BH mass estimates are $10^{8.1} M_\odot$ and $10^{7} M_\odot$ and their Eddington ratios are -1.67 / -0.72 \citep{Koss2022}. Both galaxies are part of interacting systems and are at larger distances (98.3 Mpc and 69.4 Mpc) than our ReveaLLAGN sample. Despite the increased distances, the higher luminosity results in a physically larger line-emitting region dominated by AGN photoionization, which helps to mitigate the differences in physical length scales between them and our sample. The nuclear spectra of NGC~7319 and NGC~7469 are AGN dominated and point-like at MIRI wavelengths, representing suitable examples of higher luminosity AGN with similar spectral resolution and wavelength coverage as our ReveaLLAGN targets.

In Section~\ref{sec:data} we describe the data acquisition and reduction processes. We present our spectral extraction process and emission-line measurements for both the nuclear spectra and the emission-line maps in Section~\ref{sec:methods}.  We present our analysis of the data in Section~\ref{sec:result}, and discuss them in context of previous work in Section~\ref{sec:discussion}.  We conclude in Section~\ref{sec:conclusion}.  We note that all JWST data is barycenter corrected, and thus velocities are given in the barycentric frame.

\begin{figure*}
    \centering
    \plotone{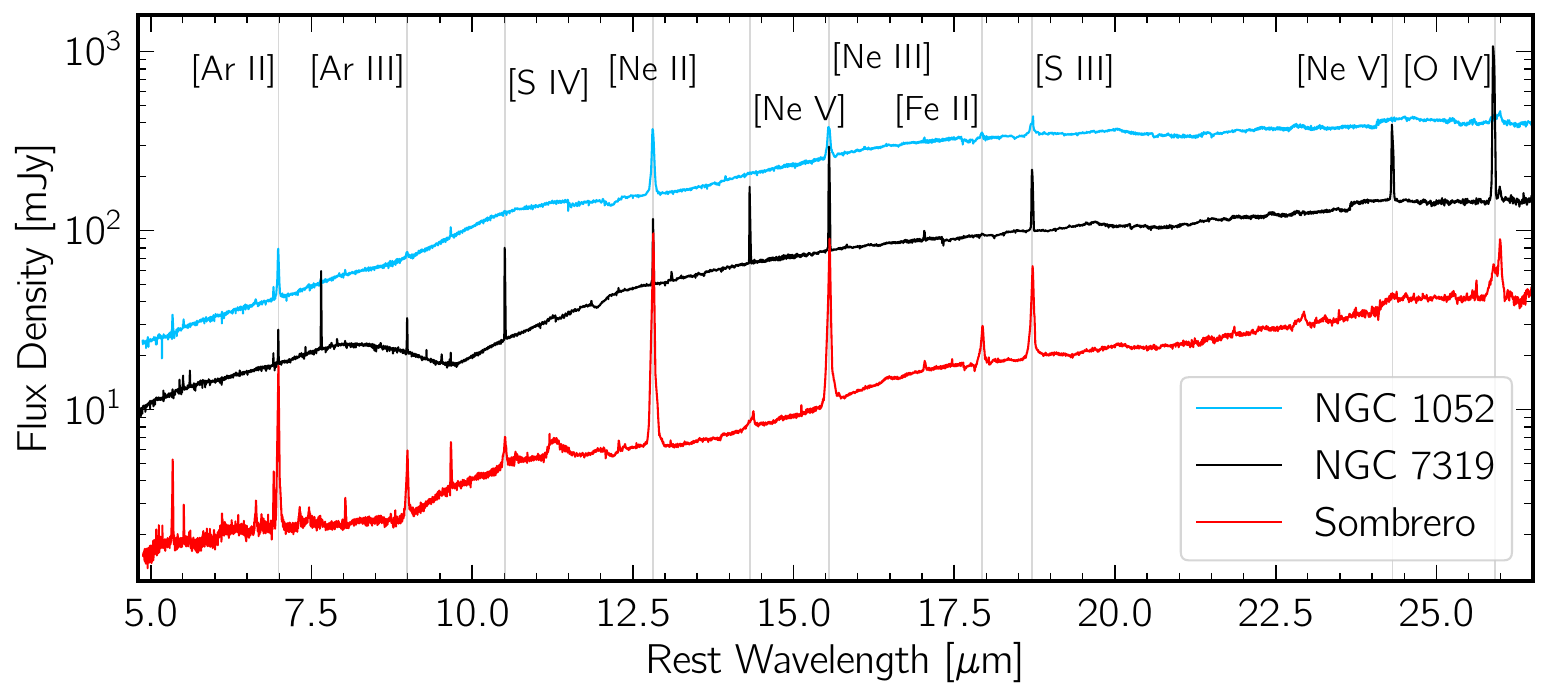}
    \caption{The first extracted nuclear spectra of ReveaLLAGN targets: Sombrero (red, bottom spectrum) and NGC~1052 (blue, top spectrum). NGC~7319 (black, middle spectrum) is a more distant and more luminous Seyfert 2 AGN and is included to compare our low-luminosity sample to another spectrum taken with JWST MIRI/MRS.  Spectra are extracted from a $\sim$1 FWHM radius aperture (see Section~\ref{sec:nuclear_extraction}) and are aperture corrected using point source observations.  A subset of strong emission lines are labeled.  Also apparent in the spectra at $\sim$10~microns are broad Silicate absorption features (in NGC~7319) and emission features (in Sombrero and NGC~1052), and faint polycyclic aromatic hydrocarbon (PAH) emission at 11.3~$\mu$m in Sombrero. }
    \label{fig:nucspec}
\end{figure*}

\begin{deluxetable*}{lcccccccc}
\tabletypesize{\footnotesize}
\setlength{\tabcolsep}{0.03in}
\tablewidth{0pt}
\tablecaption{Galaxy Properties}
\tablehead{
{Galaxy Name} & \colhead{Distance} &  \colhead{$V_{sys}$} & \colhead{Galaxy Mass} & \colhead{Morph.} & \colhead{AGN Type} & \colhead{BH Mass} & \colhead{AGN X-ray Lum.} & \colhead{Eddington Ratio}\\ 
{} & {Mpc} &  {km~s$^{-1}$} &{log(M$_\star$/M$_\odot$)} & {} & {} & {log(M$_\bullet$/M$_\odot$)} & {log($L_X$/erg s$^{-1}$)} & {log($L_{bol}/L_{edd})$}}

\startdata
NGC~1052 & 19.4$\pm$0.2 & 1487.9$\pm$5.1 & 10.71 & E4 & L1.9 & 8.82 & 41.46 & -3.97\\
Sombrero/M104/NGC~4594$^{1}$ & \phantom{1}9.6$\pm$0.3 & 1090.9$\pm$5.1 & 11.18 & Sa & L2 & 8.83 & 40.04 & -5.66 \\
\enddata
\tablerefs{ {\bf Distances}: NGC~1052 -- \citet{Tonry2001}, Sombrero -- \citet{McQuinn2016}. {\bf Systemic Velocities $V_{sys}$}: are NASA Extragalactic database heliocentric velocities taken from \citet{CORV} for NGC~1052 and \citet{RC3} for Sombrero.  {\bf Galaxy Mass}: NGC~1052 \& Sombrero from S4G \citep{Sheth2010,Eskew2012} with Sombrero corrected to the distance used here. {\bf Morphological Type}: from \citet{RC3}, {\bf AGN Type}: NGC~1052 and Sombrero from \citet{Ho1997}.  {\bf BH Mass}: NGC~1052 is based on velocity dispersion \citep{Koss2022}, Sombrero from \citet{Jardel2011}. {\bf AGN X-ray Luminosity}: 2-10~keV luminosities for NGC~1052 from \citet{Koss2022}, Sombrero from \citet{Fern2023} using updated distance.  {\bf Eddington Ratio}: NGC~1052 and Sombrero from \citet{Fern2023} using listed distances and BH masses.}
\tablenotetext{1}{We adopt ``Sombrero'' for the galaxy's name in this work.}
\label{table:galprop}
\end{deluxetable*}

\section{Data Reduction and Methods} \label{sec:data}

\subsection{Targets and Data Acquisition}

We use JWST MIRI/MRS \citep{Wells2015} to collect IFS data for our ReveaLLAGN targets in the mid-IR (4.9--27.9~$\mu$m). The full mid-IR wavelength range for MIRI/MRS is covered by 4 different channels (ch1--4):
ch1 (4.9–7.65~$\mu$m) and ch2 (7.51–11.71~$\mu$m) use the MIRIFU\_SHORT Detector, while ch3 (11.55–17.98~$\mu$m) and ch4 (17.71–27.9~$\mu$m) use the MIRIFU\_LONG Detector. Each channel has an increasing field of view (FoV): ch1 (3.2\arcsec$\times$ 3.7\arcsec), ch2 (4.0\arcsec $\times$ 4.8\arcsec), ch3 (5.2\arcsec $\times$ 6.2\arcsec), and ch4 (6.6\arcsec $\times$ 7.7\arcsec), and pixel size: ch1 (0\farcs196), ch2 (0\farcs196), ch3 (0\farcs245), ch4 (0\farcs273).  All observations were taken using all three MIRI/MRS sub-channels.

We describe the observational details for our two ReveaLLAGN targets; details on the NGC~7319 observation are discussed in \citep{Pereira2022}.  Our Sombrero observations are centered at RA: 12:39:59.430 DEC: -11:37:22.99; this is taken from Gaia EDR3 \citep{GaiaEDR3}.  Our NGC~1052 observations are centered at RA: 02:41:04.798, DEC: -08:15:20.75 taken from very-long-baseline interferometry measurements of the AGN \citep{Lambert2009}.

Background exposures were taken using offset blank fields selected based on WISE 12~$\mu$m imaging: for Sombrero this field was at RA: 12:39:55.9810, DEC: $-$11:32:11.44 and for NGC~1052 at RA: 02:41:5.1200, DEC: $-$08:12:37.70.

Our MIRI/MRS measurements were taken using the 4-Point, Extended Source optimized ALL-channel dither pattern using the inverted, or negative, dither orientation\footnote{\url{https://jwst-docs.stsci.edu/jwst-mid-infrared-instrument/miri-operations/miri-dithering/miri-mrs-dithering}}. This ensures improved sampling of the point spread function (PSF) at all wavelengths and allows the correction of hot detector pixels. The exposure time for both Sombrero and NGC~1052 was 921.313 seconds split over four dithers for each sub-channel setting. Background exposures used a single dither position with an exposure length of 230.328 seconds for each sub-channel setting.  The Sombrero data were among the first science data taken with JWST on July 4th, 2022, while the NGC1052 data were taken on August 11th, 2022. 

The JWST data presented in this paper were obtained from the Mikulski Archive for Space Telescopes (MAST) at the Space Telescope Science Institute. The specific observations analyzed can be accessed via \dataset[DOI: 10.17909/n1hq-4p52]{
https://doi.org/10.17909/n1hq-4p52
}.


\subsection{Data Reduction}\label{sec:data_reduction}

We process the raw observations for Sombrero, NGC~1052, and NGC~7319 through version 1.8.2 of the JWST pipeline \citep{bushouse2022} using jwst\_0989.pmap, which is a versioned reference file that gives overall context for the pipeline. Calibration of our data is divided into three main stages of processing; the \texttt{Detector1}, \texttt{Spec2}, and \texttt{Spec3} pipelines. 

The \texttt{Detector1} pipeline takes the raw counts from the detector, applies basic detector-level corrections to all exposures, and creates uncalibrated countrate images, or lvl2a data products\footnote{ See \href{https://jwst-docs.stsci.edu/jwst-science-calibration-pipeline-overview/stages-of-jwst-data-processing/calwebb_detector1}{calwebb\_detector1 documentation} for more information. }.  The \texttt{Spec2} pipeline takes the lvl2a products and applies additional instrumental corrections and calibrations to produce a fully calibrated individual exposure, or lvl2b data products. For MIRI/MRS observations, this stage includes adding WCS information, flat field corrections, and stray light subtraction. We include an optional fringing removal\footnote{See \href{https://jwst-docs.stsci.edu/jwst-science-calibration-pipeline-overview/stages-of-jwst-data-processing/calwebb_spec2}{calwebb\_spec2 documentation} for more information } step during this stage to address the significant fringes found in the MIRI/IFU data. The \texttt{Spec3} pipeline processes lvl2b spectroscopic observations into lvl3 data by combining calibrated lvl2b data from associated dithered exposures into a 3-D spectral cube or 2-D extracted spectra. For MIRI/MRS data the master background subtraction and outlier detection occurs in this stage as well. We choose a final product of 4 data cubes, one for each channel\footnote{See \href{https://jwst-docs.stsci.edu/jwst-science-calibration-pipeline-overview/stages-of-jwst-data-processing/calwebb_spec3}{calwebb\_spec3 documentation} for more information. }. The wavelength solution, FLT-4, associated with our pipeline version has a 1$\sigma$ wavelength calibration error of 10$-$30~km~s$^{-1}$ \citep{Arg2023} through the MRS wavelength range.

\begin{figure*}
    \centering
    \plottwo{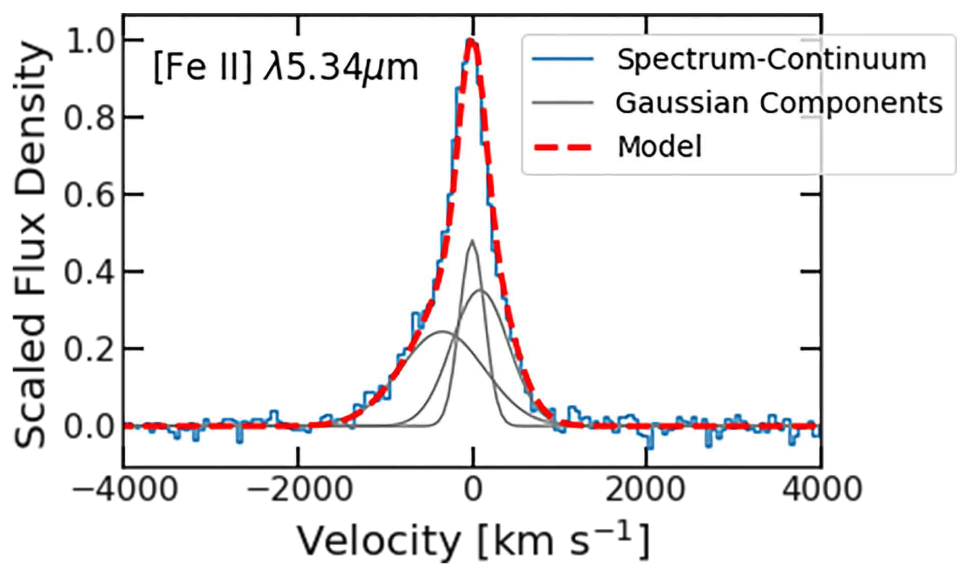}{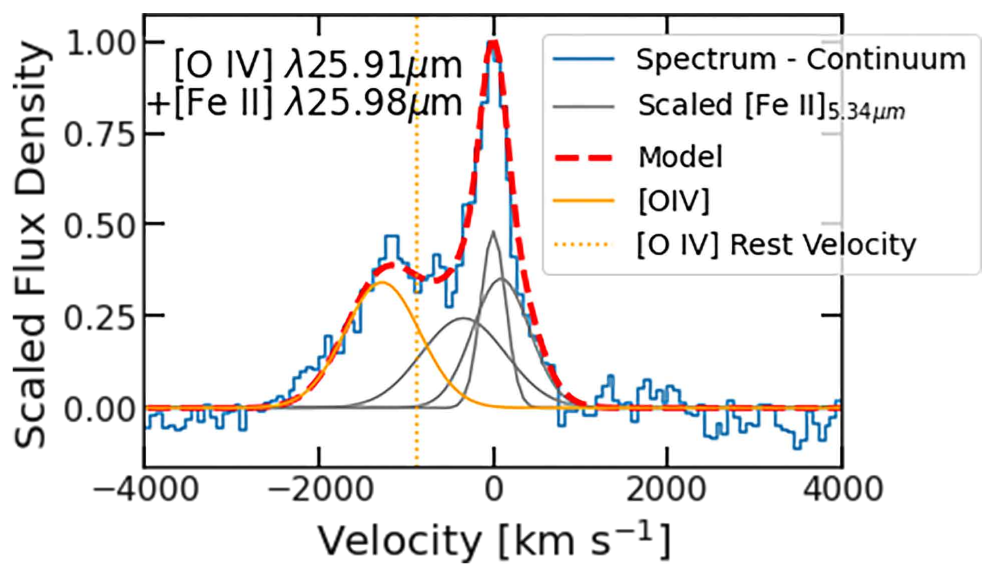}
    \caption{Two examples from Sombrero of the multi-Gaussian models used to characterize emission lines in our sample. {\em Left --} The \fetwo\ emission-line fit with a three-component Gaussian model. The gray lines represent individual Gaussian components, while the red dashed line is the sum of those three components. This is the typical method used for characterizing emission features in our data.  {\em Right -- } One of the two coronal lines requiring deblending; here we show the \ofour\ line, which is blended with [\ion{Fe}{2}]$\,\lambda$25.98$\mu$m; a scaled version of the \fetwo\ line is fit along with a single Gaussian for the [\ion{O}{4}] line.  All components are plotted relative to the velocity of [\ion{Fe}{2}]$\,\lambda$25.98$\mu$m.  Markings are as in the left panel, with [\ion{O}{4}] and its expected rest velocity shown in orange.}
    \label{fig:feii}
\end{figure*}

\section{Spectral Extraction and Methods}\label{sec:methods}

\subsection{Nuclear Spectra Extraction}\label{sec:nuclear_extraction}
Nuclear spectra were extracted using the \texttt{photutils} python package's aperture photometry code.  At each wavelength, we used a photometric aperture centroided on the median flux image of each channel.  The width of this aperture depended on wavelength to account for the changing PSF, with an angular radius of 1.22$\lambda$/(6.5~meters) -- roughly 1~spatial FWHM (${\rm FWHM}_{\rm Rayleigh}$); this aperture radius ranges from 0$\farcs$19 at 5~$\mu$m to 0$\farcs$97 at 25~$\mu$m.  The radius of this aperture on the short wavelength 0$\farcs$19 corresponds to 8.8, 17.9, and 92~pc in Sombrero, NGC~1052, and NGC~7319 respectively.  Background subtraction was done using an annulus with radii between 2 and 2.5$\times$ this value.  

We created a wavelength-dependent aperture correction based on the MIRI data cube of 10 Lac (obtained from Argyriou, I., {\em private communication}).  This aperture correction (total/aperture flux) was derived using the same aperture and background annulus as for our galaxy nuclei, with the total flux obtained by integrating the flux of the full data cube.  Due to residual sky background issues, we took the median flux of pixels with a radius greater than $6\times{\rm FWHM}_{\rm Rayleigh}$ as a background subtraction in each spaxel before calculating the total flux of the data cube at each wavelength. To create a smooth relation, we smoothed the derived aperture correction at each wavelength with a moving median. We compared this smoothed aperture correction to several other point source observations (HD192163 and HD76534) as well as NGC~1052, which is nearly point like at longer wavelengths and found generally good agreement (to within $\sim$10\%) in the aperture corrections between sources for channels 1-3, with much poorer agreement and due to noisier measurements in channel 4.  The aperture correction declines from values of $\sim$2.1 at 5\,$\mu$m to values similar to the \texttt{WebbPSF} prediction (1.4).  We therefore fit a 5th order polynomial to our smoothed correction in channels 1-3, and set the ch4 correction to a constant 1.4 value.  This aperture correction has been applied throughout this paper.

\subsection{Measuring Emission Features}\label{sec:measure_features}

\subsubsection{Multi-Gaussian Fitting of the Nuclear Spectrum}\label{sec:measure_nuclear}

Our nuclear spectra are very high S/N with clear evidence of many emission lines.  These lines often show complex profiles -- to extract both flux and velocity information from these lines, we perform multi-Gaussian fits.  We first define continuum and fitting windows for each line based on visual inspection -- our default fitting window is based on a velocity width of 5000~km$\,$s$^{-1}$.  We fit a linear function to the continuum on either side of the emission feature and subtract the result from the data. Next, we utilize the python package \texttt{lmfit} to fit both a single Gaussian and multi-Gaussian model to the continuum-subtracted emission line. We allow the multi-Gaussian model to consist of up to five components, where each Gaussian component is constrained by the width of the wavelength dependent MIRI instrument LSF and the results of the initial single-Gaussian fits. We select the model with the lowest Bayesian inference criteria (BIC) as the best-fit model. 
An example fit to \fetwo\ is shown in the left panel of Figure~\ref{fig:feii}.  We do not ascribe any physical interpretation to the individual Gaussian components, instead, we use them to accurately describe the emission-line profile from which we measure the flux, peak velocity, and ${\rm FWHM}_{\rm model}$. The ${\rm FWHM}_{\rm line}$ of each emission-line is corrected for the width of the MIRI/MRS line spread function (LSF) at the corresponding wavelength, given by 
\begin{equation}\label{eqn:fwhmline}
{\rm FWHM_{\rm line}} = \sqrt{{\rm FWHM}_{\rm model}^{2} - {\rm FWHM}_{\rm LSF}^{2}}
\end{equation}
We use the MIRI MRS LSF width given by \citet{Arg2023}: ${\rm FWHM}_{\rm LSF} = c/R$, where $c$ is the speed of light, and $R =$ 4603 - 128$\lambda$.

Errors on derived quantities are determined from a Monte Carlo (MC) simulation with Gaussian noise added to each pixel based on the standard deviation of the pixels in the continuum windows.  The median standard deviation in the continuum pixels is $\sim$4$\times$ the formal flux errors provided by the pipeline.  Emission-line detections are determined if the integrated flux of the best single-Gaussian emission-line model is above a 5$\sigma$ threshold. 5$\sigma$ upper limits are provided for lines without clear detections.  We adopt a lower limit on errors for any wavelength dependent measurement equal to the wavelength calibration error of 30 km~s$^{-1}$ provided in \ref{sec:data_reduction}. The derived line properties and their associated errors are given in Table~\ref{tab:values}.

Two key lines of interest for tracing AGN activity are the high-IP lines (IP $>$ 50 eV) \nefive\ and \ofour.  
However in both our ReveaLLAGN targets, these lines are each blended with a neighboring low-IP line (IP $<20)$.  Specifically, \nefive\ is blended with the [\ion{Cl}{2}]$\,\lambda$14.36\,$\mu$m emission line, while \ofour\ is blended with the [\ion{Fe}{2}]$\,\lambda$25.98\,$\mu$m emission line. We deblend the features using a constrained multi-Gaussian model; the low-IP component is fixed to be a scaled version of the [\ion{Fe}{2}]$\,\lambda$5.34\,$\mu$m line (Figure~\ref{fig:feii}), an isolated low-IP line with high signal-to-noise.  We then allow \texttt{lmfit} to fit the \nefive\ and \ofour\ emission with a single Gaussian component. To capture the full uncertainty of this measurement we fit the [\ion{Fe}{2}]$\,\lambda$5.34\,$\mu$m in each iteration of the MC process before constraining the \nefive\ and \ofour\ models.

\subsubsection{Constructing Emission Line Maps}\label{sec:measure_linemap}

Outside the nucleus, many lines have low signal-to-noise ratios, making the multi-Gaussian method we use for the nuclear spectrum less robust.  We therefore simplify the Gaussian fitting process used for the nuclear spectra described above by limiting the Gaussian model to a single Gaussian component.  The emission-line flux is calculated by measuring the area under the best-fit Gaussian model, while velocity is determined by calculating the displacement between the centroid of the best-fit Gaussian model and the rest wavelength of the emission line. For the blended high-IP features (e.g.~Fig.~\ref{fig:feii}, right), we attempted to deblend them pixel-by-pixel using two-Gaussian fits, but found no significant detection of the \nefive\ and \ofour\ emission beyond the central few spaxels due to a combination of low S/N and perhaps the nuclear concentration of these lines.  We calculate errors on the flux and velocity using a Monte Carlo simulation as above, and use a 5$\sigma$ detection threshold, below which we find our Gaussian fits don't characterize the data well.  We discuss the resulting line maps in the Section~\ref{subsec:linemaps}. 

To investigate the ionizing mechanisms of our emission lines, we quantify the spatial extent of the emission region in our line maps by measuring the spatial FWHM (${\rm FWHM}_{\rm spat}$) of prominent emission lines. We do this by creating a contour at 50\% of the peak flux and calculate 2$\times$ the median radius from the peak flux to the contour line. 
We correct the measured ${\rm FWHM}_{\rm spat}$ for the MIRI/MRS PSF, which varies by a factor of five over the MIRI wavelength range. Using the FWHM of the MIRI/MRS PSF (FWHM$_{\rm MRS}$) taken from \cite{Arg2023} we get: 
\begin{equation}\label{eqn:fwhmspat}
{\rm FWHM_{\rm spat,corr}} = \sqrt{{\rm FWHM}_{\rm spat}^{2} - {\rm FWHM}_{\rm MRS}^{2}}
\end{equation}
The results for this measurement are listed in Table \ref{tab:FWHM} and presented in Section~\ref{sec:FWHM}, with discussion in \ref{sec:lit_review}.

\section{Results}\label{sec:result}

\subsection{Nuclear Region Emission Line Analysis}\label{sec:nuclear}

\subsubsection{Variations with Ionization Potential }\label{sec:ipvar}

\begin{figure}
    \center
    \includegraphics[width=\linewidth]{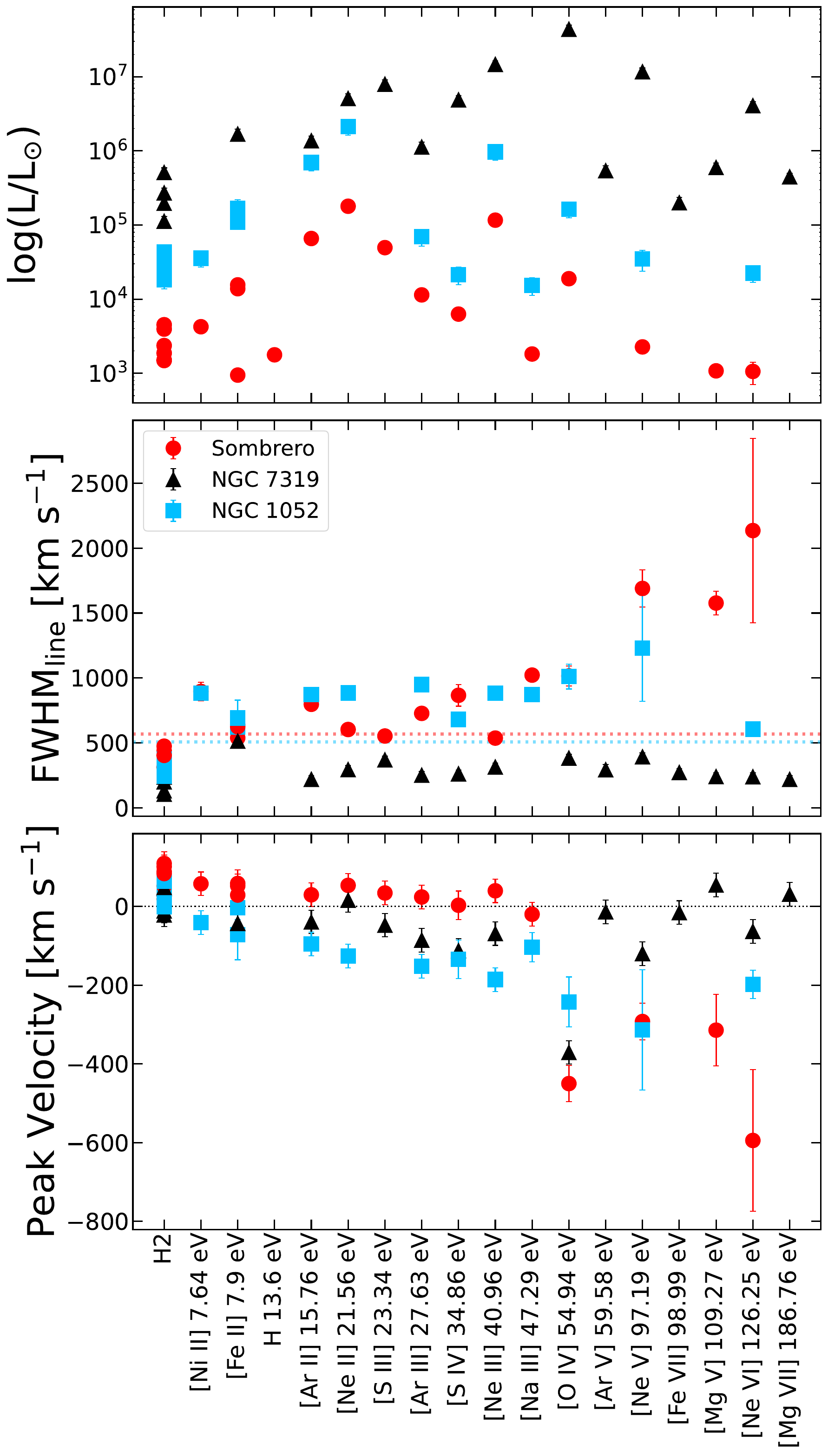}
    \caption{Emission-line trends with ionization potential.  Emission features are listed along the x-axis ordered by their IP.  \emph{Top -- Luminosity vs IP}. Emission-line luminosities scale with the Eddington ratio of sources. NGC~7319 has the highest Eddington ratio and the most luminous emission lines, followed by NGC~1052, and then Sombrero. The luminosities have a median fractional error of 15\%.
    \emph{Middle -- FWHM$_{line}$ vs IP}. The FWHM$_{\rm line}$ of emission features increases with IP in Sombrero and NGC~1052 while the FWHM$_{\rm line}$ of NGC~7319 emission features stays relatively constant with IP. FWHM$_{\rm line}$ in km~s$^{-1}$ is shown on the y-axis with a median error of 30 km~s$^{-1}$. Red and blue dashed lines represent the central stellar velocity dispersion measurements from \citet{Ho2009b} translated to a FWHM. 
    \emph{Bottom -- Peak Velocity vs IP}. Peak velocity of emission lines trend increasingly blue-shifted with increasing ionization potential in Sombrero and NGC~1052. The y-axis shows the peak velocity of the best fit Guassian model with a median error of 30 km~s$^{-1}$. }
    \label{fig:lum-fwhm-vel}
\end{figure}

In Figure~\ref{fig:lum-fwhm-vel} we show the nuclear emission-line properties in our two ReveaLLAGN targets, as well as NGC~7319, ordered by their IP to search for systematic trends.  The top panel shows the line luminosity and we find the most luminous detected lines in Sombrero and NGC~1052 are \netwo\ followed by \nethree\ which have IPs of 21.56 and 40.96 eV respectively, while in NGC~7319 the \ofour\ line (IP$=$54.94~eV) is the most luminous line.  More generally, NGC~7319 shows overall higher luminosity  in all lines compared to Sombrero and NGC~1052, with the relative luminosity increasing for the higher IP lines.

The middle panel of Figure~\ref{fig:lum-fwhm-vel} shows the ${\rm FWHM}_{\rm line}$ (see equation \ref{eqn:fwhmline}) of each line as a function of IP.  These ${\rm FWHM}_{\rm line}$ values are derived from the best-fit multi-Gaussian model to the nuclear emission lines (Section~\ref{sec:measure_nuclear}). The red and blue dashed lines represent arcsecond-level central velocity dispersions for Sombrero and NGC~1052 \citep{Ho2009b} translated to a FWHM. The emission lines in Sombrero and NGC~1052 are visibly broader than those in NGC~7319 (as can be seen in Figure \ref{fig:nucspec}).  Specifically, in NGC~7319 the lines have ${\rm FWHM}_{\rm line}\sim$200~km$\,$s$^{-1}$ regardless of IP. Meanwhile in Sombrero and NGC~1052, all detected lines are significantly wider, with the broadest lines having ${\rm FWHM}_{\rm line}\gtrsim$1000~km$\,$s$^{-1}$.  A clear trend is also seen with IP in Sombrero with the higher IP lines having significantly larger ${\rm FWHM}_{\rm line}$ values. A similar trend is seen in NGC~1052 though with the [\ion{Ne}{6}]$\,\lambda$7.65$\mu$m emission feature being notably narrower than other high-IP lines. A similar correlation is found between FWHM and IP in NGC~7469 \citep{Armus2023} when comparing the FWHM of the broad components of the emission lines. The widths of these components range from approximately 600 km s$^{-1}$ to 1100 km s$^{-1}$, falling between the ranges seen in NGC~7319 and Sombrero.

Finally, the bottom panel of Figure~\ref{fig:lum-fwhm-vel} shows the peak velocity of the emission lines as a function of IP.  The peak velocity is measured from our best-fit multi-Gaussian models and we see distinct differences between the galaxies here.  For NGC~7319, the peak velocities are quite close to zero at all IP, which some slightly blue-shifted lines ($\sim$50~km$\,$s$^{-1}$) at intermediate IPs.  The exception is the [\ion{O}{4}] line, which shows a significant blue-shift. We caution that this line is one of the longest wavelength lines we have; the wavelength calibration is less accurate at long wavelengths, but is still estimated to be $<$30~km$\,$s$^{-1}$ by \citet{Arg2023}; this line is also among the most blue-shifted lines in Sombrero and NGC~1052. 

For Sombrero, the high-IP lines are almost all significantly blueshifted (greater than 3$\sigma$ from zero), while the lower IP lines and H$_2$ lines show a slight redshift.  The redshift of the H$_2$ lines in Sombrero (median Peak Velocity of 56 km~s$^{-1}$) may indicate that our systemic velocity taken from HI measurements \citep{RC3} is offset; if this were the case most of the low- and mid-IP lines would show a modest blue-shift with a general trend of larger blue-shift with higher IP.  In NGC~1052, the blueshift in the highest IP lines are weaker, but there is also a sign of blue-shifted emission even at lower IP. The blue-shifted emission could be due to outflows, which we discuss in detail in Section~\ref{sec:outflow_discussion}.

\subsubsection{Detailed Nuclear Line Profiles}

\begin{figure}
    \center
    \includegraphics[width=\linewidth]{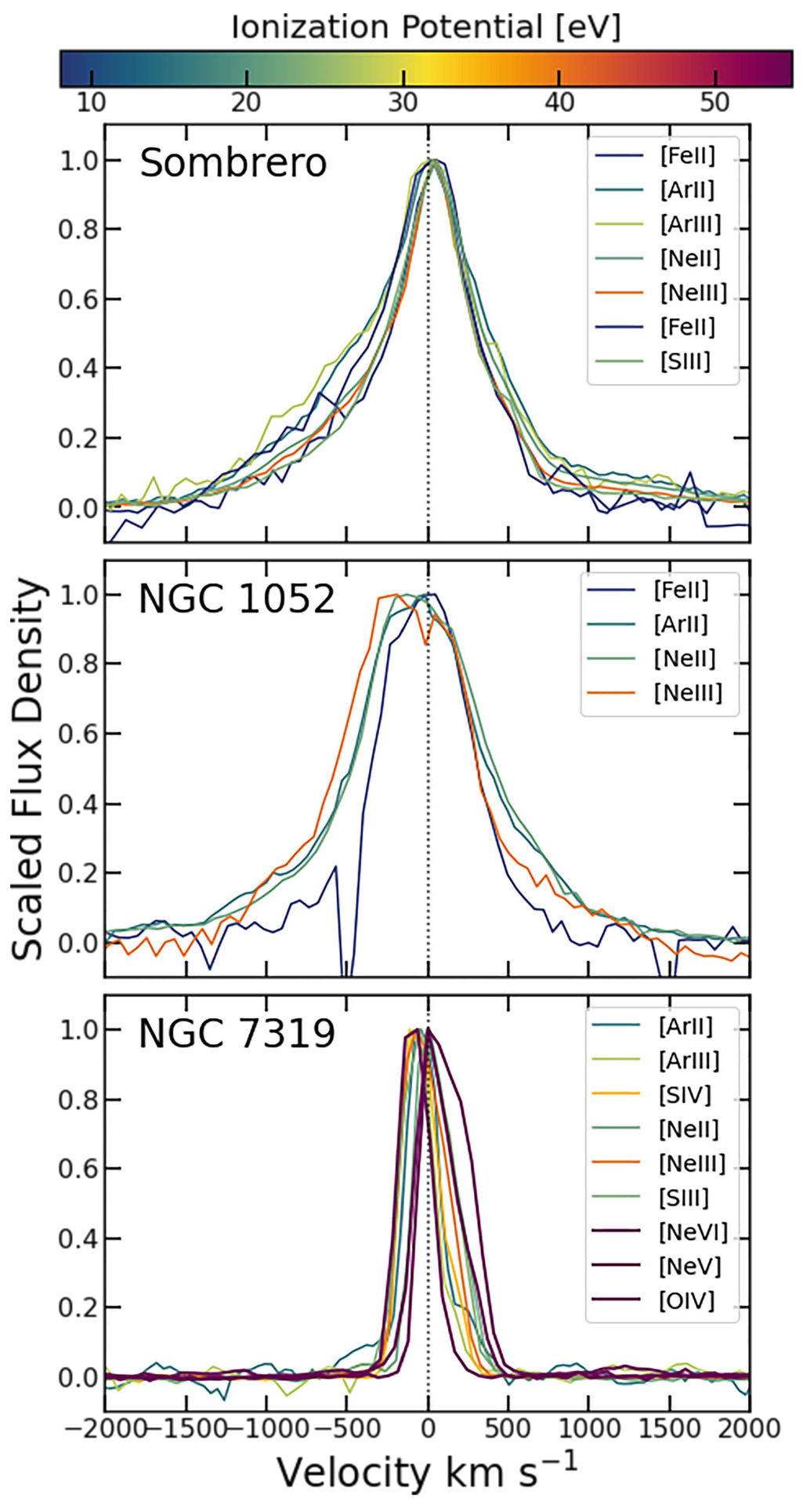}
    \caption{Nuclear emission-line profiles with ${\rm S/N} >50$ centered on expected velocity. Sombrero lines are asymmetrical with a blueshifted extension, or wing, while NGC~1052 and NGC~7319 have generally more symmetric profiles with blue-shifted peaks. Emission lines in NGC~7319 show red-shifted extensions at high IP.}
    \label{fig:lineprofiles}
\end{figure}

The high spectral resolution of JWST lets us resolve line widths and look at the detailed shapes of emission lines.  Above we found that the high-IP lines show broad, often blue-shifted emission lines, and here we look in more detail at the shapes of the lines with the highest signal-to-noise ratios (S/N $>$ 50).  Figure~\ref{fig:lineprofiles} shows these lines in each galaxy centered on their expected velocity. 
Looking at each galaxy, these strong lines show remarkably consistent line profiles suggesting a common physical origin.  However, significant differences are seen between galaxies, with Sombrero having a notably asymmetric line profile with blue wings reaching $>$1000~km$\,$s$^{-1}$, while NGC~1052 and NGC~7319 show more symmetric lines.  The strong asymmetry in Sombrero likely indicates the presence of an outflow, which we will discuss in more detail in Section~\ref{sec:outflow_discussion}. Blue asymmetries are also observed in the highest IP emission line profiles of NGC~7469 \citep{Armus2023}.  The narrower lines in NGC~7319 relative to the other two galaxies are clearly visible as well.  We note that the highest IP lines in NGC~1052 and Sombrero are not high enough S/N to examine their line profiles in detail (as well as blending issues in a couple lines) .

\subsection{2-D Emission Line Information: Line Maps \& FWHM}\label{sec:maps}

\begin{figure*}
    \center
    \includegraphics[width=\textwidth, height=.49\textheight,,keepaspectratio]{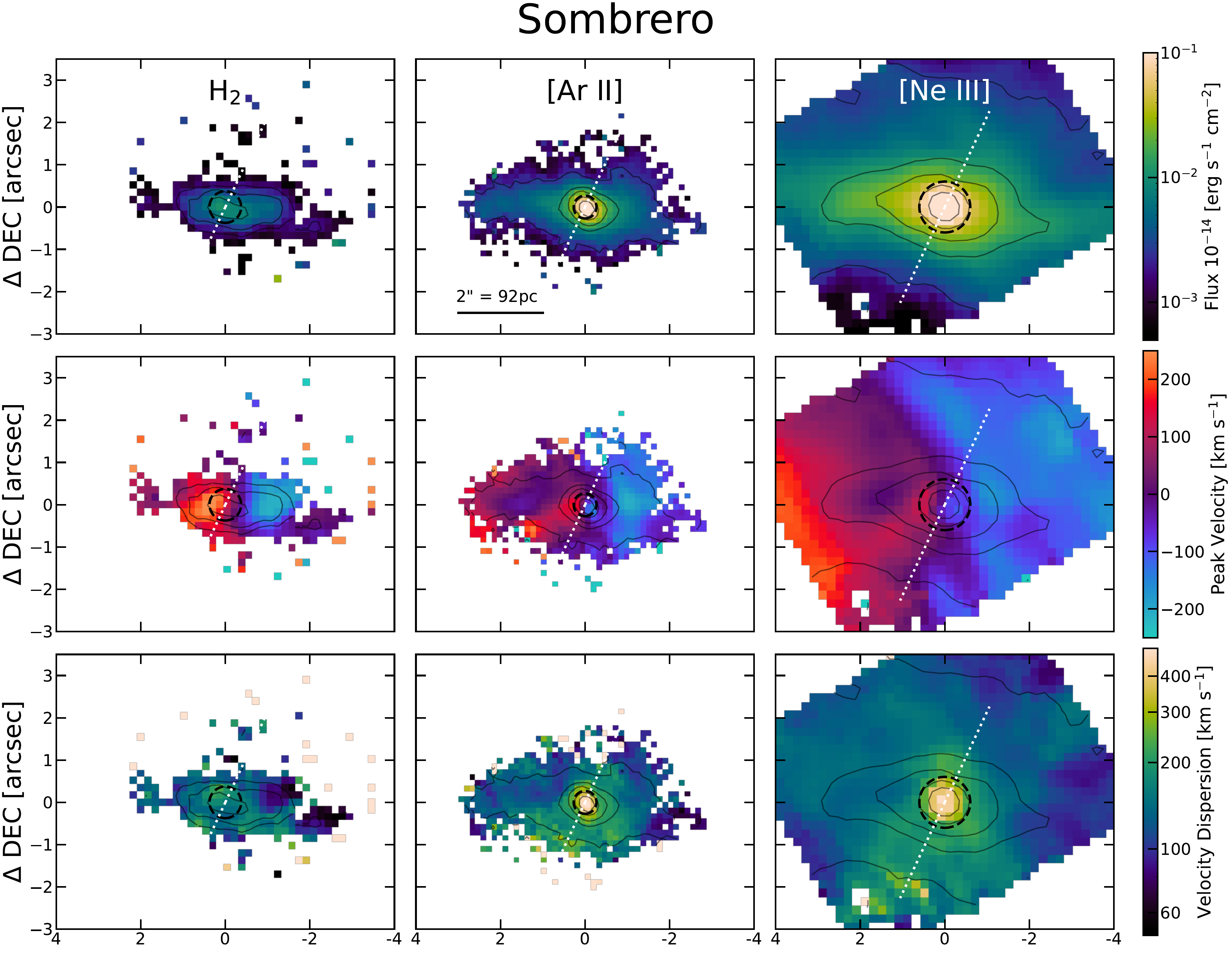}
     \includegraphics[width=\textwidth, height=.49\textheight,,keepaspectratio]{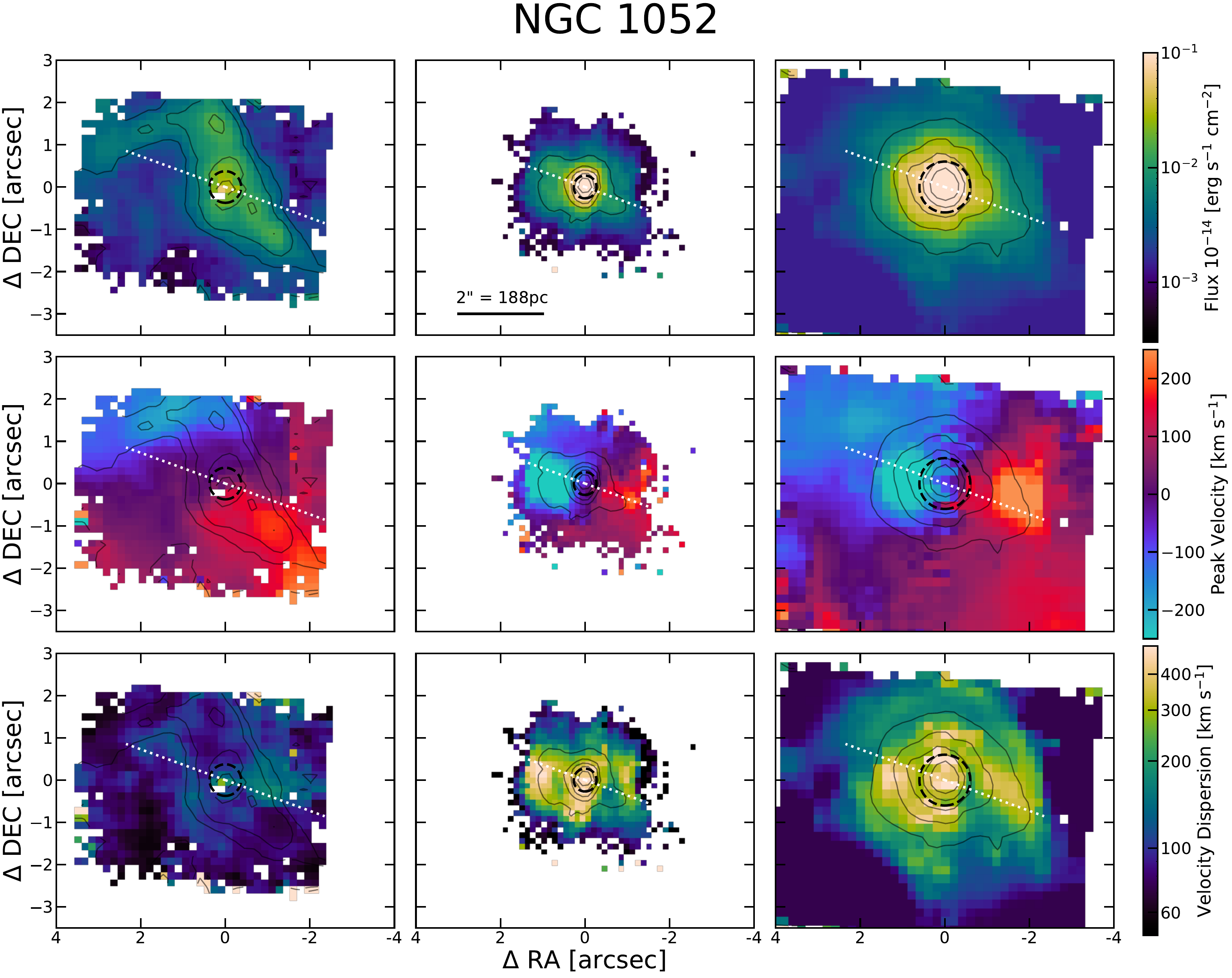}
    \caption{Flux, velocity, and dispersion maps for three emission lines in both Sombrero and NGC~1052. In all maps north is up and east is to the left. The leftmost column shows the H$_2$(0-0)S(3) molecular hydrogen line at 9.66\,$\mu$m, the middle column shows the low-IP line [\ion{Ar}{2}]$\,\lambda$6.98\,$\mu$m, and the right column shows the mid-IP line \nethree.  Contours indicate flux levels of 1, 5, 10, 25, and 50\% of the peak line flux, while the dashed black line represents the aperture used to extract the nuclear spectrum at that wavelength. The white dotted lines, (shown with arbitrary length) indicate the orientation of compact radio jets; corresponding to a PA of $-25^\circ$ in Sombrero oriented nearly along our line-of-sight \citep{Hada2013}, and a PA of 70$^\circ$ in NGC~1052 oriented along the plane of the sky \citep{Kadler2004}.}
    
    \label{fig:linemaps}
\end{figure*}

\subsubsection{Flux and Velocity Maps}
\label{subsec:linemaps}

Figure \ref{fig:linemaps} shows flux and velocity maps for three lines in both Sombrero and NGC~1052. 
These are created using the single Gaussian fitting method described in Section~\ref{sec:measure_linemap}.  Three lines are shown for each galaxy; The H$_{2}$ 0–0 S(3) line at 9.66~$\mu$m, the [\ion{Ar}{2}] line at 6.98 $\mu$m (IP: 15.76~eV), and the [\ion{Ne}{3}] line at 15.56~$\mu$m (IP: 40.96 eV).  These three lines span a wide range of IP and critical densities and thus likely trace very different density gas \citep[e.g.][]{Stern2014}.  The highest IP lines (IP $>$ 50~eV) are unresolved, and therefore compact, showing detectable emission only in the central few pixels.   

In the Sombrero galaxy, all three lines have similar morphologies, extended east-to-west with blue-shifted emission towards the west.  The molecular hydrogen emission has no clear point-like emission and is red-shifted relative to the systemic velocity in the nuclear region; this redshift is also seen in several other H$_2$ and low IP lines in Sombrero(Figure~\ref{fig:lum-fwhm-vel}). As discussed in the previous subsection, this may be due to the adopted systemic velocity for Sombrero. The velocity dispersion seen in molecular hydrogen emission maps is quite homogeneous with values up to 240 km~s$^{-1}$, comparable to the measured nuclear stellar velocity dispersion \citep[241 km~s$^{-1}$;][]{Ho2009b}. Clear point-like emission is seen in both [\ion{Ar}{2}]$\,\lambda$6.98$\mu$m and \nethree;  this emission appears to be more concentrated in [\ion{Ar}{2}]$\,\lambda$6.98$\mu$m than \nethree, however this may be due simply to the lower resolution at these wavelengths; we examine this in more detail below in Section~\ref{sec:FWHM}.  Filaments can be seen extending out to the north/west from the nuclear region in the \nethree\ flux map. The velocity maps of both ions shown are similar to H$_2$ (red-shifted to the east, and blue-shifted to the west), but show complex velocity fields e.g.~a patch of blue-shifted emission $\sim$2\arcsec~east of the nucleus and a stretch of red-shifted emission stretching south-east from the nuclear region . The velocity dispersion in [\ion{Ar}{2}]$\,\lambda$6.98$\mu$m and \nethree\ both peak in the nuclear region with a maximum velocity of about 500 km~s$^{-1}$.

In NGC~1052, the H$_2$ emission-line map differs significantly from the [\ion{Ar}{2}]$\,\lambda$6.98$\mu$m  and \nethree\ emission. The H$_2$ emission-line flux maps have a weak peak in the nuclear region and extend north-east to south-west. The velocity maps of H$_2$ are blue-shifted in the north-east and red-shifted to the south and west. The velocity dispersion is larger along the minor axis of rotation and peaks at $\sim$275 km~s$^{-1}$ in the nuclear region, a bit higher than the \citet{Ho2009b} central stellar velocity dispersion of 215 km~s$^-1$. The H$_2$ flux, velocity, and dispersion maps presented here for NGC~1052 are in agreement with \cite{muller2013} where the H$_2$ 1-0 S(1) line at 2.12$\micron$ was examined using SINFONI, benefiting from slightly better spatial resolution. \cite{muller2013} interpret the morphology and kinematics of H$_2$ as a decoupled rotating disk, due to the gas having a kinematic major axis that is not aligned with the stellar rotation axis. Our H$_2$ flux map is also similar in morphology to the CO gas seen with ALMA in \citep{Kameno2020}, which they interpret as a circumnuclear disk. The [\ion{Ar}{2}]$\,\lambda$6.98$\mu$m  and \nethree\ emission-line flux maps are strongly peaked in the nucleus and share a roughly concentric radial profile. The corresponding velocity maps of NGC~1052 reveal extended emission with a distinct kinematic structure characterized by a heavily blue-shifted region directly East of the nucleus and a heavily red-shifted region to the West, with velocities up to 590 km~s$^{-1}$. As detailed in Section~\ref{sec:intro}, NGC~1052 has an inner radio jet on $\sim$2~pc scales with a PA of $\sim$70 degrees \citep{claussen1998, Kadler2004}, while at larger scales ($\sim$1 kpc) the PA of the radio jets is approximately 100 degrees \citep{Wrobel1984, Kadler2004a}. Our MIRI/MRS data falls between these two scales, and the PA of the kinematic structure we see (Figure~\ref{fig:linemaps}) falls between the PAs of these inner and outer jets.

\subsubsection{Spatial FWHM Measurements}
\label{sec:FWHM}

Following the methodology outlined in Section~\ref{subsec:linemaps}, we determine FWHM$_{\rm spat,corr}$, characterizing the PSF-corrected spatial extent, for six emission lines in Sombrero and four emission lines in NGC~1052.  These lines are at low- and mid- IP and have sufficient signal-to-noise to enable the measurement. The FWHM$_{\rm MRS}$, FWHM$_{\rm spat}$ and FWHM$_{\rm spat,corr}$ measurements are provided in Table \ref{tab:FWHM}. 
Overall, we find that the lines in NGC~1052 are either unresolved or just barely spatially resolved, with the [\ion{Ne}{3}] line having the largest spatial extent (FWHM$_{\rm spat,corr} = 0\farcs30$ or 28.2~pc).  On the other hand, all the emission lines in Sombrero are spatially resolved, with FWHM$_{\rm spat,corr} > 0\farcs17$ or 8~pc, and no clear trend with IP.  We note that while FWHM$_{\rm spat,corr}$ estimates were not possible for the high-IP coronal lines (\ofour\ and \nefive), these lines do appear to be quite compact in both galaxies.  In both galaxies, the \nethree\ emission is more extended than the \netwo\ emission, a somewhat surprising result that we discuss further in Section~\ref{sec:lit_review}.

\section{Discussion}\label{sec:discussion}
In this section we present our results in the context of previous work. First, in section \ref{sec:resolving}, we discuss the power of JWST in separating LLAGN from their host galaxies. Then in section \ref{sec:lit_review}, we compare the nuclear emission features from our LLAGN to AGNs of varying types, and end with section \ref{sec:outflow_discussion} by discussing evidence for outflows seen in the LLAGN spectra.

\begin{figure}
    \center
    \includegraphics[width=\linewidth]{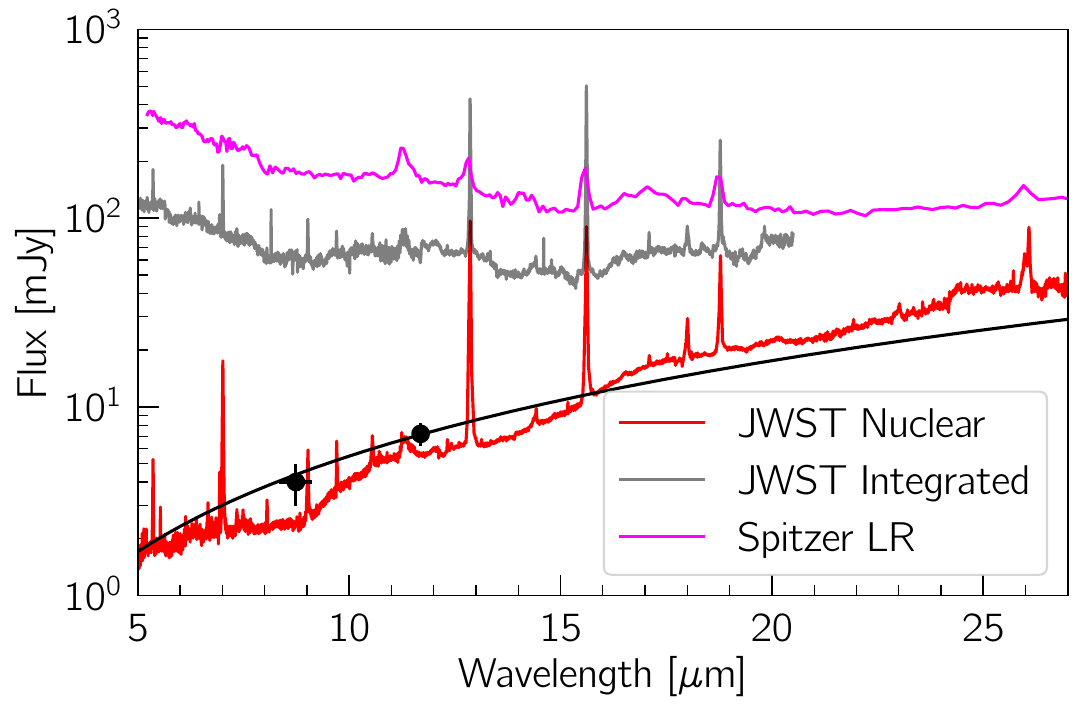}
    \caption{JWST enables us to separate LLAGN spectra from their host galaxy. 
 Comparison of the aperture corrected nuclear extracted spectrum in Sombrero (red line; same as Figure~\ref{fig:nucspec}), to the integrated MIRI/MRS spectrum (gray line; FoV: 6$\farcs$6$\times$7$\farcs$7),and the Spitzer LR spectrum  (magenta line; FoV: 27$\farcs$7$\times$51$\farcs$8).  The black line shows the best fit high-spatial resolution power-law fit to Sombrero from \citet{Fern2023} this is fit to the black points, which are photometry from Gemini \citep{asmus2014} and VLT \citep{Fern2023} as well as  sub-arcsecond data at shorter wavelengths; both the data and fit are in good agreement with our nuclear spectrum. We show the integrated spectrum only out to 20~$\mu$m as the poorly constrained MIRI channel 4 background levels significantly impact the integrated spectrum measurements at redder wavelengths. }
    \label{fig:spitzercomp}
\end{figure}

\subsection{The Promise of JWST for Revealing LLAGN}
\label{sec:resolving}

In Figure~\ref{fig:spitzercomp} we show a comparison of the extracted nuclear spectrum (see Section \ref{sec:nuclear_extraction}) in Sombrero to both the integrated flux in the JWST data cube, and the Spitzer LR spectrum from the SINGS survey \citep{Kennicutt2003}.  The integrated flux was calculated by summing all spaxels in each MIRI data cube. Since the FoV varies between each channel, we normalized the integrated spectrum to channel 4. In this channel the FoV measures $6\farcs6\times7\farcs7$ corresponding to a physical scale of $306\times357$~pc$^2$ at the distance of Sombrero.  Note that the integrated spectrum is not shown at the longest wavelengths due to sky subtraction issues as discussed in \citet{Gasman2023}.  

The nuclear emission clearly shows a SED that increases with wavelength, while the integrated data cube has a very different SED.  Just $\sim$1\% of the flux in the JWST integrated cube is coming from the nuclear component at 5\,$\mu$m,  while the nuclear compoment is $>$10\% of the flux by 20\,$\mu$m.  This rising nuclear SED is consistent with two previous photometric measurements of Sombrero at high resolution (black points/line in Figure~\ref{fig:spitzercomp}) and within the expectations of LLAGN spectra \citep{Fern2023}. However, the information available in the nuclear spectrum is clearly far richer than was available with previous ground-based photometric measurements.  

The two larger scale spectra from both Spitzer and our integrated JWST data in Figure~\ref{fig:spitzercomp} show very different spectral shapes that are dominated by galaxy emission. The shape of these two spectra are in good agreement despite the different apertures suggesting a roughly constant SED for the galaxy component.  Overall, the data show that even in Sombrero, the faintest target in the ReveaLLAGN survey, we can cleanly extract the LLAGN emission and separate it from its surrounding galaxy.
Although the primary goal of this paper is analysis of the emission lines in our ReveaLLAGN MIRI spectra, the continuum shape also encodes information on the emission mechanisms of these LLAGN.  High angular resolution work on LLAGN has consistently shown jet dominated emission to follow a broken power-law continuum \citep{Ho1996, Chary2000, Prieto2016,Fern2023} which is consistent with self-absorbed synchrotron emission characteristic of compact jet emission \citep{Marscher1985}.

While Figure~\ref{fig:spitzercomp} shows broad agreement with a single power-law fit from \citet{Fern2023} over the MIRI wavelength range, there is also considerable complexity seen in the SEDs (Figure~\ref{fig:nucspec}), with a clear inflection point in the Sombrero nuclear spectrum at ~9 $\mu$m.  We also see a gradual flattening of the spectrum at long wavelengths in NGC~1052, which is consistent with the turnover of the broken power below ~20 µm and the nuclear fluxes at lower frequencies \citep{Fern2019}.  The complexity of the continuum shapes we see in the MIRI spectra suggest additional information may be available from detailed fitting of the continuum that includes the contributions of broad silicate features (Fern\'andez-Ontiveros et al., {\em in prep}).

\subsection{The Emission Lines of LLAGN: Comparison to Previous Work}\label{sec:lit_review}

\begin{figure}
    \center
    \includegraphics[width=\linewidth]{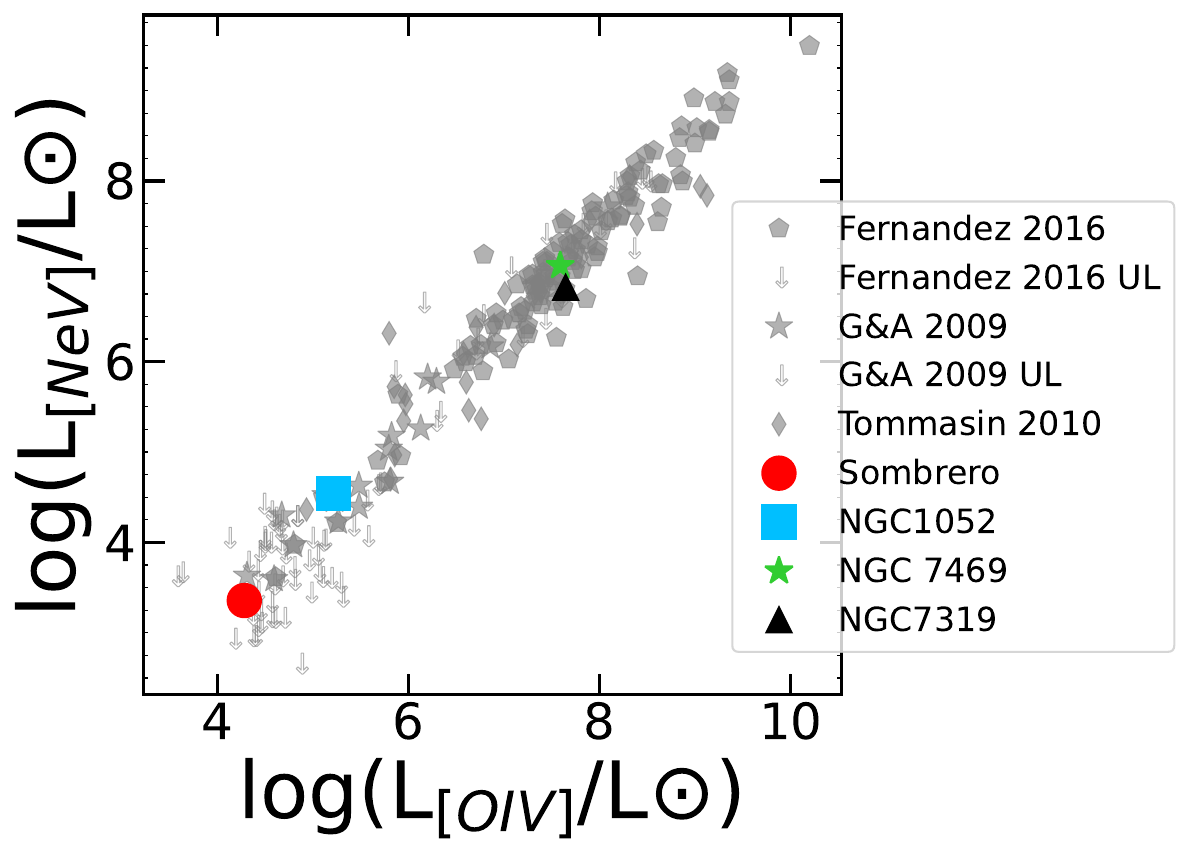}
    \caption{Sombrero and NGC~1052 have very low luminosity detections of \nefive\ and \ofour, 
    with \nefive\ in Sombrero having one of the lowest luminosity detections to date. A tight, nearly linear relationship can be seen when comparing the luminosities of coronoal lines \ofour\ (x-axis) and \nefive\ (y-axis). The logarithm of the luminosity on both axes is shown in solar units (3.846$\times10^{33}{\rm erg}\ {\rm s}^{-1}$).  Gray markers represent results from previous surveys \citep{Goulding2009,Tommasin2010,Fernandez2016} with upper limits on [\ion{Ne}{5}] found in \cite{Goulding2009} and \cite{Fernandez2016}. The green star represents measurements for NGC~7469 taken from \cite{Armus2023}} 
    \label{fig:nev_oiv}
\end{figure}

In this subsection, we focus on comparing the nuclear emission-line luminosities and ratios to previous measurements of typically much higher luminosity AGN.   

Figure \ref{fig:nev_oiv} compares the luminosities of the two high-IP lines detected in all three galaxies, \nefive\ and \ofour\ to literature measurements primarily from Spitzer \citep{ Goulding2009, Tommasin2010, Fernandez2016}.
We note that these data have much lower physical resolution than our nuclear JWST data, and thus contamination of the AGN spectra by galaxy light is likely significant in some cases, especially for lower-IP lines discussed below that are excited by sources other than the AGN.  NGC~7319 and NGC~7469, as expected, have luminosities very typical of previously measured AGN, while Sombrero has the lowest luminosities of both lines compared to any previous measurements. While Sombrero and NGC~1052 stand out as being very low luminosity detections, they both follow the tight, nearly linear correlation between these two coronal lines that is seen across a wide range of AGNs \citep{Goulding2009}.

\begin{figure*}
    \center
    \includegraphics[width=\linewidth]{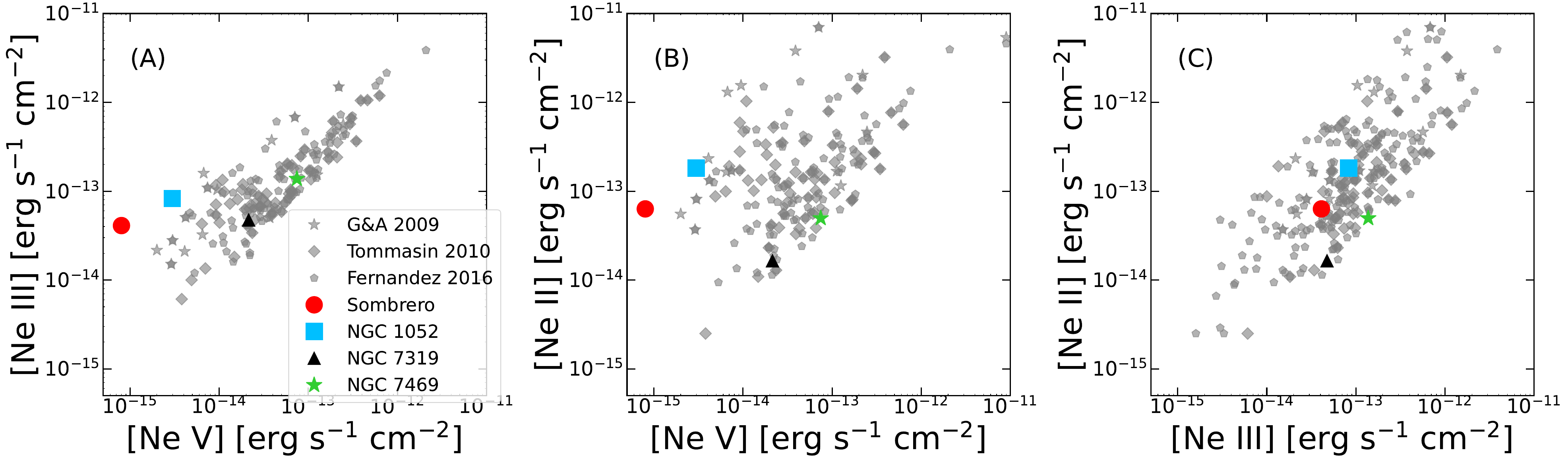}
    \caption{Flux measurements of different ionized states of neon from our sample compared to previous AGN surveys. The \nefive\ flux from our sample, especially in Sombero, is much lower than in previously observed AGN relative to the \nethree\  (Plot A) and \netwo\ (Plot B) fluxes.  However, the [\ion{Ne}{2}] and [\ion{Ne}{3}] fluxes are fairly typical of other AGN (plot C).  
    Units on all axes are in erg $\mathrm{s}^{-1}\,\mathrm{cm}^{-2}$. Markers are the same as in \ref{fig:nev_oiv}.}
    \label{fig:nev_neiii_neii}
\end{figure*}

Comparing ionized states of a particular atom enables us to study the ionization structure within an AGN more clearly. In this regard, the mid-IR is particularly valuable as it contains multiple neon emission lines at different ionizations. In Figure \ref{fig:nev_neiii_neii} we compare the flux values of \netwo, \nethree, and \nefive\ from our sample to previous surveys. Comparing line fluxes (rather than luminosities) ensures that correlations seen are the result of excitation differences, and not caused by observing sources at a range of distances (which can create false correlations between line luminosities).

The left panel comparing [\ion{Ne}{5}]\footnote{For the rest of the discussion, we will refer to \netwo, \nethree\ and \nefive\ as [\ion{Ne}{2}], [\ion{Ne}{3}] and [\ion{Ne}{5}], respectively.} and [\ion{Ne}{3}] shows a roughly linear correlation that gets tighter with increasing [\ion{Ne}{5}] flux.  Sombrero has significantly weaker [\ion{Ne}{5}] than other sources with similar [\ion{Ne}{3}] flux, and many of the lower luminosity sources including NGC~1052 also scatter towards fainter [\ion{Ne}{5}] flux relative to the relation seen at higher line fluxes.  Thus Sombrero is an outlier, but follows the qualitative trend of lower [\ion{Ne}{5}] luminosity that are seen in other lower luminosity AGN.  The middle panel comparing the flux of [\ion{Ne}{2}] to [\ion{Ne}{5}] shows similar results to the left panel, but with a much looser relation seen between the lines at high line fluxes.  Finally the right panel shows that the relative [\ion{Ne}{2}] and [\ion{Ne}{3}] flux fall within the range of previous measurements in all three galaxies.  This suggests that these lower IP lines have values typical of higher luminosity AGN, and it is the [\ion{Ne}{5}] line that is weaker than in other sources.

\begin{figure}
    \center
    \includegraphics[width=\linewidth]{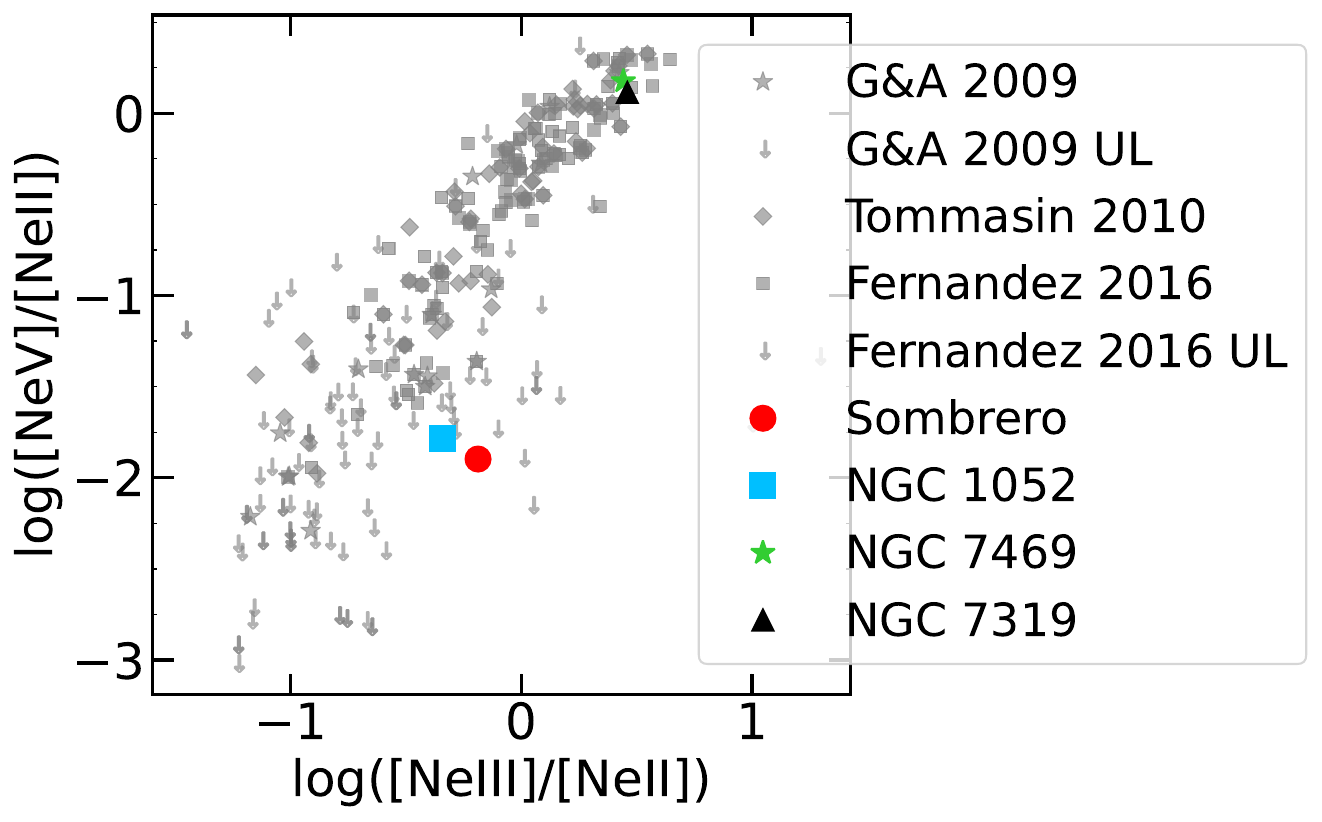}
    \caption{The low-luminosty detections of [\ion{Ne}{5}] place our LLAGN sample well below the trend line when comparing the logarithm of  \nethree/\netwo\ (x-axis) to the logarithm of \nefive/\netwo\ (y-axis). Marker colors are the same as in Figure \ref{fig:nev_oiv}, and \ref{fig:nev_neiii_neii}.}
    \label{fig:contribution}
\end{figure}

We combine the information on all three neon lines in Figure~\ref{fig:contribution}, which compares the ratios of [\ion{Ne}{5}]/[\ion{Ne}{2}] and [\ion{Ne}{3}]/[\ion{Ne}{2}].  The ratio of [\ion{Ne}{5}] to [\ion{Ne}{2}] has been employed as a diagnostic tool in IR spectra to assess the contribution of AGN activity \citep{Goulding2009, Sajina2022}. Since [\ion{Ne}{5}] can only be formed through AGN processes, while [\ion{Ne}{2}] can arise from both AGN and non-AGN mechanisms, this ratio helps determine the presence and influence of AGN. We emphasize again, that the literature data here have low spatial resolution, and therefore any line emission in the central kiloparsecs of the galaxies contain significant contamination from the host galaxy. 
 NGC~1052 and especially Sombrero fall well below the main trend line found in Figure \ref{fig:contribution} and into a region only populated with upper limits of [\ion{Ne}{5}] from other surveys.

We can get a sense of the level of galaxy contamination in our own JWST spectra by comparing the extent of emission features with different IP and in Section \ref{sec:FWHM} we find that the ${\rm FWHM}_{\rm spat,corr}$ of the [\ion{Ne}{2}] and [\ion{Ne}{3}] emission lines are quite compact.  We would expect [\ion{Ne}{2}] be more spatially extended than higher IP lines, including [\ion{Ne}{3}], since [\ion{Ne}{2}] lines come predominantly from star formation.  This is not what we find in either source; in fact [\ion{Ne}{2}] is found to be more compact than [\ion{Ne}{3}] in both NGC~1052 and Sombrero.  The fact that [\ion{Ne}{2}] emission is compact doesn't strictly mean that it comes from the AGN, it could simply mean that any star formation is also compact/unresolved.  While \cite{Prieto2014} reports the presence of extended H$\alpha$ emission perpendicular to the jet in Sombrero, which may be associated with star formation, they find no conclusive evidence of star formation, from UV to IR, within parsecs of the center of Sombrero, nor in NGC~1052 \citep{Prieto2021}.  A lack of excitation from star formation is consistent with the absence of any PAH emission in  the nuclear spectra of NGC~1052 and only a weak PAH signature at 11.3 $\mu$m in Sombrero (Fig.~\ref{fig:nucspec}).   This lack of evidence for star formation suggests that the nuclear line ratios from our targets (Figure~\ref{fig:contribution}) are not significantly contaminated by emission from star formation, and that the outlier status of our two galaxies are the result of very low luminosity detections of [\ion{Ne}{5}] made possible by the spatial and spectral resolution of JWST.  The differences we see then in Figure~\ref{fig:nev_neiii_neii} are due to excitation differences from the AGN accretion structure.  This difference can be explained by either a change in SED or very low ionization parameters that result in a deficiency of the high energy photons ($\gtrsim$100~eV) needed to excite the line.  This conclusion is consistent with previous work on LLAGNs \citep{Ho2008, Eracleous2010} including photoionisation models for compact jet synchrotron emission \citep{Fern2023}, shock excitation models \citep{Dopita2015}, and the expectations of a central engine with advection dominated accretion flows \citep{Nemmen2014}. We will be able to test this result and compare this to models for AGN ionization once the full ReveaLLAGN sample is available (Fern\'andez-Ontiveros et al., {\em in prep}).

\subsection{Outflows in NGC~1052 and Sombrero}\label{sec:outflow_discussion}
In Section~\ref{sec:nuclear}, we identify the following emission-line features in NGC~1052 and Sombrero:

\vspace{-2mm}
\begin{itemize}
    \setlength\itemsep{-1.5mm}
    \item an increase in line widths with IP
    \item an increase in blue-shifted emission with IP
    \item broad emission in the weakly-detected high-IP and coronal lines, and
    \item prominent blue wings in the high signal-to-noise lines of Sombrero.
\end{itemize}\vspace{-2mm}

The trend of increasing line width and IP was originally attributed to cloud stratification--the coronal lines are emitted from denser clouds closer to the central engine which are subject to more intense ionizing flux \citep{filippenko1984,filippenko1985,filippenko1988,appenzeller1988}. Recent work has confirmed that many Seyfert galaxies, regardless of brightness or AGN type, show an increase in both line FWHM and line blue-shifting with increasing IP \citep[e.g.,][]{rodriguez2006,rodriguez2011,Armus2023}. Furthermore, there are known correlations between blue-shifted emission and both increasing IP in coronal lines and increasing line width in the [\ion{O}{3}] line in narrow-line Seyfert~1 galaxies \citep[e.g.][]{Komossa2008}. While there is clear evidence that coronal-line emission and their profiles are driven mainly by photoionization from the AGN \citep[e.g.,][]{Nussbaumer1970,Korista1989,Oliva1994,Pier1995,rodriguez2011}, other work has demonstrated that outflows are needed to fully explain the observed emission \citep[e.g.,][]{appenzeller1988,Erkens1997,wilson1999,rodriguez2006,rodriguez2011,muller2011}. In fact, the blue-shifted emission even at mid-IPs could trace out-flowing material closer to the AGN than the narrower emission, with the line asymmetry being caused by red-shifted emission being absorbed along the line-of-sight \citep{Komossa2008}.

Given the known importance of outflows and shocked emission in LINERs \citep[e.g.,][]{Ho2008, Trump2011, Molina2018}, we conclude that the emission-line features identified above are indicators of outflows for both Sombrero and NGC~1052. We discuss other evidence and the possible origins of the outflows in NGC~1052 and Sombrero below.

\subsubsection{Previous Evidence of Outflows in NGC~1052}\label{ssec:1052_outflow}

Previous work has demonstrated the presence of AGN-related outflows in NGC~1052 on multiple spatial scales. Optical IFS studies of NGC~1052 show evidence for an outflow from the AGN on larger scales \citep{Sugai2005, Dopita2015, Dahmer-Hahn2019,Cazzoli2022}. The outflow is roughly aligned with the radio jet \citep{claussen1998,Kadler2004}, with a PA of $\sim$70$^\circ$ and is generally in good agreement with the velocity structures seen in Figure~\ref{fig:linemaps}. These studies also find a broad H$\alpha$ and H$\beta$ component with a width of $\sim$3000~km$\,$s$^{-1}$; this is significantly broader than the widths of the mid and high-IP lines we see here.

Similarly, on much smaller spatial scales, \cite{muller2013} finds evidence of outflows in the velocity dispersion maps of H$_2$ emission seen in the IR, while \cite{Pogge2000}, \cite{Walsh2008} and \cite{Molina2018} found evidence for outflows in \textit{HST} data. Both \cite{Pogge2000} and \cite{Walsh2008} found evidence for strong outflows as well as ionized regions associated with jet-like features. Meanwhile, \cite{Molina2018} demonstrated that shocked emission likely originating from these outflows are the dominant power source at just $\sim$20~pc outside of the galaxy center. Similar to \cite{Dopita2015}, \cite{Cazzoli2022} and this work, \cite{Molina2018} found that the shock-dominated, off-nuclear emission lines had widths consistent with $v\lesssim500$~km~s$^{-1}$. They also found broad H$\alpha$ and H$\beta$ emission in the unresolved AGN spectrum, with ${\rm FWHM}\sim10^3$~km~s$^{-1}$. We note that a majority of the emission seen in \cite{Molina2018} lies within the JWST nuclear aperture used in this work. 

\subsubsection{Previous Evidence of Outflows in Sombrero}\label{ssec:sombrero_outflow}

Given the low accretion rate and the presence of a small-scale radio jet, Sombrero likely has strong radio outflows \citep{Meier2001,Fender2004}. In fact, \cite{Walsh2008} determined that while Sombrero has organized motion within the central 0\farcs5 consistent with an overall rotation pattern, there are significant irregularities that could be caused by outflows. \cite{Pogge2000} also found evidence of turbulent motion via spiral-like wisps in the narrow-band H$\alpha$+[\ion{N}{2}] imaging. \cite{emsellem2000} further identified a strong velocity gradient near the galaxy center, and noted that the kinematics of the gas within the central 1\arcsec\ was decoupled from the gas in the spiral wisps. These east-west oriented wisps are not well-aligned with the inner radio jet described by \cite{Hada2013} and \cite{Mezcua2014}, which runs along the north-south axis and is oriented towards our line of sight. We note that the presence of broad H$\alpha$ is unclear, with  two analyses of the same HST spectra coming to different conclusions \citep{Walsh2008, hermosa2020}.  \cite{Mason2015} found that the near-infrared SED appears to be similar to that of other type 2 LINERs, and \cite{Gallimore2006} and \cite{Li2011} also found evidence for larger-scale outflows in Sombrero using radio and X-ray data, respectively.

\subsubsection{Origins of Outflows}\label{ssec:outflow_origin}

Here we consider two possible models for the outflows seen in NGC~1052 and Sombrero. We note that radiation pressure-driven outflows do not significantly contribute to the outflows seen in LLAGN \citep{Meena2023}, and therefore we do not discuss them below. As a reminder, both of these objects are classified as LINERs and exhibit low Eddington ratios (Table~\ref{table:galprop}), with evidence of compact radio jets (Section~\ref{sec:intro}; Section~\ref{subsec:linemaps}). 

\vspace*{2mm}\noindent{\it Winds Launched from the RIAFs:}

Unlike traditional cold, thin-disk models, RIAFs occur when the accretion rate is sufficiently low that the inner disk puffs up and becomes a hot, advection-dominated accretion flow \citep{Narayan1995,Blandford1999, Yuan2014}. Previous empirical studies showed that radio outflows from AGNs, including those with thin-disk accretion flows and RIAFs, increase in strength as the accretion rate decreases \citep[e.g.,][]{Ho2002,Melendez2010}. RIAFs extending to large scales can eliminate broad line emission \citep{Elitzur2009} and the ``big blue bump'' associated with thin-disk accretion \citep[e.g.,][]{Trump2011}; the corresponding lack of UV emission and broad line features in most LINER AGN \citep[e.g.,][]{Nicastro2003,Ho2008} suggests they may be powered by RIAFs. 

The strong wind along the polar or jet direction in RIAFs that was predicted by magnetohydrodynamical numeral simulations \citep{Yuan2012,Yuan2015} has been observationally confirmed in recent years \citep[e.g.][]{Wang2013,Cheung2016,Park2019,Shi2021}. These energetic winds originate in the coronal region of the accretion flow, implying that higher-IP lines would experience more intense outflows, and thus likely have larger widths, consistent with the findings presented in Section~\ref{sec:nuclear}. Given their low accretion rates (see Table~\ref{table:galprop}), the absence of the ``big blue bump'' in both of their SEDs \citep{fernandez2012}, and the lack of clear broad H$\alpha$ emission in Sombrero \citep{Walsh2008}, it is likely that both NGC~1052 and Sombrero are powered by a RIAF. Therefore, we conclude that the energetic winds driven by the hot accretion flows in both LLAGNs likely contribute to the observed emission. However, we note that by their nature RIAFs do drive radio jets, and as such these winds may not be the sole explanation for the observed outflows.

\vspace*{2mm}\noindent{\it Jet-Driven Outflows:}

Jets associated with AGN accretion are known to drive outflows that create shocked emission and can regulate the star-formation rate in the galaxy \citep[e.g.,][]{Silk1998,Weinberger2017,dave2019}. In fact, while we did not find any trends with IP in the nuclear spectra of NGC~7319, \cite{Pereira2022} found that high-IP coronal-line emission is detected close to the hot spots of the known radio jet, which they conclude indicates the presence of a jet-driven outflow.

Due to their less luminous, lower-accretion rate engines, the shocked emission driven by jets or outflows can often dominate over photoionization at small distances from the nuclei in LLAGNs \citep{Molina2018}. Furthermore subparsec-scale radio jets occur more frequently in LINERs \citep{Nagar2005}, which could further indicate the presence of jet-driven outflows.

Recent work by \cite{Meenakshi2022} demonstrated that small-scale jets can produce large widths even in mid-IP lines like [\ion{O}{3}]\,$\lambda$5007, similar to the widths seen in our mid-IP lines studied here. They also conclude that similar widths can be seen in the different gas phases of the ISM, which appears to be somewhat qualitatively true for NGC~1052--the observed positive correlation between IP and FWHM in NGC~1052 in Figure~\ref{fig:lum-fwhm-vel} is much less pronounced than that in Sombrero. Furthermore, both \cite{Sugai2005} and \cite{Dopita2015} found evidence that the jet in NGC~1052 was interacting with the circumnuclear gas. 

In both the RIAF- and jet-driven wind scenario, the orientation of the jet should impact the observable signatures.  In Sombrero, modeling of VLBI data suggests the inner jet is oriented close to our line-of-sight \citep{Hada2013}, while in NGC~1052, the jet is oriented more in the plane of the sky \citep{Kadler2004}.  This difference in jet orientation may be the reason that only Sombrero shows the blue-shifted emission in its nuclear spectrum, while the ionized emission-line maps in NGC~1052 show strong strong blue- and red-shifts oriented close to the jet axis (Figure~\ref{fig:linemaps}). However, since both RIAF- and jet-driven winds will result in an outflow in the jet direction, a combination of SED modeling on the smallest scales with emission-line analysis like that presented here is likely required to resolve what drives the outflows in LLAGN.

\section{Conclusions}\label{sec:conclusion}

This paper features the first observations of the ReveaLLAGN survey, a JWST project to characterize seven nearby LLAGN.  We present MIRI/MRS data of the least and most luminous targets in our sample, Sombrero and NGC~1052.   We compare this data to that of higher luminosity AGNs, specifically NGC~7319 and NGC~4395.  We characterize the numerous emission lines seen in the nuclear spectrum
and create line maps across the MRS field of view for stronger lines.  

We find the following results:
\begin{itemize}
\item The resolution and sensitivity of JWST allows us to cleanly separate the AGN continuum and emission lines from the surrounding galaxy even in our least luminous target, Sombrero.  
\item The ionized emission lines in both Sombrero and NGC~1052 are broad, and have widths that increase with increasing IP reaching FWHM$>$1000km$\,$s$^{-1}$.  The highest IP lines (IP $>$50) show blue-shifted peak velocities with a median velocity of $-$423 km~s$^{-1}$ seen in Sombrero and $-$186 km~s$^{-1}$ in NGC~1052.
\item The highest signal-to-noise ionic lines in Sombrero with  show a clear blue wing extending $>$1000km$\,$s$^{-1}$ from the peak emission.  
\item Sombrero has the lowest luminosity high-IP lines ([\ion{O}{4}] and [\ion{Ne}{5}]) yet detected in any source.  NGC~1052 also shows low luminosity in both these lines, and the relative luminosity of these lines follows the relation seen in more luminous AGN.
\item The \nefive\ is weak relative to the \netwo\ and \nethree\ as compared to previously measured AGN.  This does not appear to be due to galaxy contamination, and thus likely indicates a deficiency of high energy ionizing photons in these LLAGN.
\end{itemize}

Our full ReveaLLAGN dataset will include observations of seven nearby LLAGN with both the NIRSpec IFU and MIRI/MRS.  We will present the nuclear spectra of these in an upcoming paper (Seth et al., {\em in prep}), as well as an analysis of their emission lines (Goold et al. {\em in prep}).  We will also be modeling the continuum emission and emission lines from the ReveaLLAGN sample (Fern\'andez-Ontiveros et al. {\em in prep}).   The ReveaLLAGN spectra will be valuable in both identifying the unique features of LLAGN, and revealing the nature of the central engine in LLAGN.    

\begin{acknowledgments}
We thank Ioannis Argyriou for his helpful suggestions and willingness to share data and the anonymous referee for their useful comments that helped improve the paper.  KG, AS, and DO acknowledge support from JWST Cycle 1 grant GO-2016. We acknowledge the ERO team for developing their observing program with a zero-exclusive-access period. The work of MM is supported in part through a fellowship sponsored by the Willard L. Eccles Foundation. LCH was supported by the National Science Foundation of China (11721303, 11991052, 12011540375, 12233001), the National Key R\&D Program of China (2022YFF0503401), and the China Manned Space Project (CMS-CSST-2021-A04, CMS-CSST-2021-A06).\end{acknowledgments}

%

\vspace{5mm}
\facilities{JWST (MIRI/MRS)}


\software{astropy \citep{Whelan2018}, lmfit (\url{https://github.com/lmfit/lmfit-py}), jwst calibration pipeline v1.8.2 (\url{https://github.com/spacetelescope/jwst})}

\movetabledown=5cm
\begin{rotatetable}
\begin{deluxetable*}{cccccccccccccc}
\tabletypesize{\footnotesize}
\setlength{\tabcolsep}{0.05in}
\tablewidth{0pt}
\tablecaption{Nuclear Spectra Measurements \label{tab:values}}
\tablehead{
\colhead{Galaxy} & \colhead{Line} & \colhead{Wavelength\tablenotemark{a}} & \colhead{IP\tablenotemark{b}} & \colhead{Transition} & \colhead{Flux} & \colhead{Flux Err}  & \colhead{Peak Vel} & \colhead{Peak Vel Err} &\colhead{ FWHM$_{\rm line}$} & \colhead{FWHM$_{\rm line}$ Err} & \colhead{S/N} & \colhead{Warning}\\
\colhead{ } & \colhead{ } & \colhead{$\mathrm{\mu m}$} & \colhead{$\mathrm{eV}$} &  \colhead{} &\colhead{$10^{-14}\mathrm{erg\,s^{-1}\,cm^{-2}}$} & \colhead{$10^{-14}\mathrm{erg\,s^{-1}\,cm^{-2}}$ } & \colhead{$\mathrm{km\,s^{-1}}$} & \colhead{$\mathrm{km\,s^{-1}}$}  & \colhead{$\mathrm{km\,s^{-1}}$} & \colhead{$\mathrm{km\,s^{-1}}$} & \colhead{} & \colhead{}} 
\startdata
Sombrero & [Fe II] & 5.340 & 7.90 & $^{4}$F$_{\frac{9}{2}}$-a $^{6}$D$_{\frac{9}{2}}$ & 0.488 & 0.004 & 50 & 30 & 540 & 30 & 94.5 & 0 \\
Sombrero & H$_{2}$ & 5.448 & 15.37 & (12-10)O(9) & $<$ 0.009 & -- & -- & -- & -- & -- & 0.0 & 0 \\
Sombrero & [Mg VII] & 5.504 & 186.76 & $^{3}$P$_{2}$-$^{3}$P$_{1}$ & $<$0.020 & -- & -- & -- & -- & -- & 1.7 & 0 \\
Sombrero & H$_{2}$ & 5.511 & 15.37 & (0-0)S(7) & 0.083 & 0.003 & 110 & 30 & 310 & 30 & 25.8 & 0 \\
Sombrero & [Mg V] & 5.608 & 109.27 & $^{3}$P$_{1}$-$^{3}$P$_{2}$ & 0.038 & 0.004 & -310 & 100 & 1580 & 110 & 7.9 & 0 \\
Sombrero & H$_{2}$ & 6.109 & 15.37 & (0-0)S(6) & 0.053 & 0.005 & 20 & 30 & 320 & 160 & 12.5 & 0 \\
Sombrero & [Ni II] & 6.636 & 7.64 & $^{2}$ D$_{\frac{3}{2}}$-$^{2}$ D$_{\frac{5}{2}}$ & 0.150 & 0.004 & 60 & 30 & 900 & 80 & 35.2 & 0 \\
Sombrero & [Fe II] & 6.721 & 7.90 & $^{4}$F$_{\frac{9}{2}}$-a $^{6}$D$_{\frac{7}{2}}$ & 0.033 & 0.003 & 60 & 30 & 620 & 50 & 9.5 & 0 \\
Sombrero & H$_{2}$ & 6.909 & 15.37 & (0-0)S(5) & 0.159 & 0.003 & 100 & 30 & 400 & 30 & 49.9 & 0 \\
Sombrero & [Ar II] & 6.985 & 15.76 & $^{2}$ P${\frac{1}{2}}$-$^{2}$ P${\frac{3}{2}}$ & 2.320 & 0.007 & 30 & 30 & 800 & 30 & 370.5 & 0 \\
Sombrero & [Na III] & 7.318 & 47.29 & $^{2}$ P$_{\frac{1}{2}}$-$^{2}$ P$_{\frac{3}{2}}$ & 0.064 & 0.003 & -20 & 30 & 1020 & 40 & 20.8 & 0 \\
Sombrero & H & 7.458 & 13.60 & Pfund-alpha & 0.063 & 0.003 & 130 & 30 & 1130 & 30 & 18.5 & 0 \\
Sombrero & [Ne VI] & 7.652 & 126.25 & $^{2}$P$_{\frac{3}{2}}$-$^{2}$P$_{\frac{1}{2}}$ & 0.037 & 0.010 & -590 & 170 & 2140 & 550 & 6.7 & 0 \\
Sombrero & H$_{2}$ & 8.026 & 15.37 & (0-0)S(4) & 0.053 & 0.001 & 50 & 30 & 440 & 30 & 32.7 & 0 \\
Sombrero & [Ar III] & 8.991 & 27.63 & $^{3}$P$_{1}$-$^{3}$P$_{2}$ & 0.403 & 0.007 & 20 & 30 & 730 & 50 & 89.3 & 0 \\
Sombrero & [Fe VII] & 9.527 & 98.99 & $^{3}$F$_{3}$-$^{3}$F$_{2}$ & $<$2.195 & -- & -- & -- & -- & -- & 0.1 & 0 \\
Sombrero & H$_{2}$ & 9.665 & 15.37 & (0-0)S(3) & 0.140 & 0.002 & 90 & 30 & 400 & 30 & 58.0 & 0 \\
Sombrero & [S IV] & 10.510 & 34.86 & $^{2}$P$_{\frac{3}{2}}$-$^{2}$P$_{\frac{1}{2}}$ & 0.222 & 0.006 & 0 & 40 & 870 & 100 & 31.8 & 0 \\
Sombrero & H$_{2}$ & 12.278 & 15.37 & (0-0)S(2) & 0.042 & 0.002 & 60 & 30 & 530 & 40 & 16.6 & 1 \\
Sombrero & H & 12.367 & 13.60 & Humph-alpha & 0.050 & 0.004 & -130 & 80 & 1670 & 190 & 15.7 & 1 \\
Sombrero & [Ne II] & 12.814 & 21.56 & $^{2}$P$_{\frac{1}{2}}$-$^{2}$P$_{\frac{3}{2}}$ & 6.317 & 0.020 & 50 & 30 & 600 & 30 & 757.2 & 0 \\
Sombrero & [Ar V] & 13.102 & 59.58 & $^{3}$P$_{1}$-$^{3}$P$_{0}$& $<$0.004 & -- & -- & -- & -- & -- & 4.0 & 0 \\
Sombrero & [Ne V] & 14.322 & 97.19 & $^{3}$P$_{}$-$^{3}$P$_{1}$& 0.080 & 0.004 & -290 & 40 & 1690 & 140 & 32.7 & 0 \\
Sombrero & [Cl II] & 14.368 & 12.97 & $^{3}$P$_{1}$-$^{3}$P$_{2}$& -- & -- & -- & -- & -- & -- & 13.3 & 3 \\
Sombrero & [Ne III] & 15.555 & 40.96 & $^{3}$P$_{1}$-$^{3}$P$_{2}$& 4.101 & 0.015 & 40 & 30 & 540 & 30 & 556.2 & 0 \\
\enddata

\tablecomments{The complete table is presented in the online version of the {\it Astrophysical Journal}. Here we present the first few rows to show its form and content. The measured quantities provided here are derived from the multi-component Gaussian fits described in Section~\ref{sec:measure_nuclear}. We define the line as detected if the integrated flux of a best-fit single-Gaussian model has a ${\rm S/N}\geq 5$; upper limits are provided for undetected emission lines. The ``Warning'' column identifies issues with the spectra (blended feature, bad pixel, etc). 0 $-$ good fit; measurements reported. 1 $-$ blended/possibly blended features based on visual inspection; measurements reported. 2 $-$ unacceptable spectra quality; no measurements to report. 3 $-$ no measurements to report due to deblending procedure (Section \ref{sec:measure_nuclear}, Figure \ref{fig:feii}). \\}
\tablenotetext{a}{Rest wavelengths from \href{https://physics.nist.gov/PhysRefData/ASD/lines_form.html}{NIST.}}
\tablenotetext{b}{Ionization potential energy from \href{https://physics.nist.gov/PhysRefData/ASD/ionEnergy.html}{NIST.}}
\end{deluxetable*}
\end{rotatetable}

\clearpage

\begin{deluxetable*}{c c c c c |c c c |c c}[h]
\tabletypesize{\footnotesize}
\setlength{\tabcolsep}{0.05in}
\tablewidth{0pt}
\tablecaption{Spatial FWHM Measurements of the Resolved Emission Lines\label{tab:FWHM}}
\tablehead{
\colhead{Feature} & \colhead{Rest Wavelength} & \colhead{IP} & \colhead{FWHM$_{\rm MRS}$} & \multicolumn{3}{c}{Sombrero}  & \multicolumn{3}{c}{NGC~1052} \vspace{-2mm}\\
 \cmidrule(l{2pt}r{2pt}){5-7} \cmidrule(l{2pt}r{2pt}){8-10}
 \multicolumn{4}{c}{} & {FWHM$_{\rm spat}$} &\multicolumn{2}{c}{FWHM$_{\rm spat,corr}$} & {FWHM$_{\rm spat}$} & \multicolumn{2}{c}{FWHM$_{\rm spat,corr}$}\vspace{-2mm}\\
 & {($\mu$m)} & {eV} & {(arcsec)}  & {(arcsec)} &{(arcsec)} & {(pc)} & {(arcsec)} & {(arcsec)} & {(pc)}}

\startdata
[Fe II] & 5.34 & 7.9 & 0.27 & 0.49 & 0.42 & 19.45 & 0.36 & 0.24 & 22.34  \\\relax
[Ar II] & 6.99 & 15.76 & 0.31 & 0.35 & 0.17 & 7.87 & 0.33 & 0.12 & 10.35  \\\relax
[Ar III] & 8.99 & 27.63 & 0.42 & 0.46 & 0.20 & 9.26 & 0.41 & --$^{\ddag}$ & --  \\\relax
[Ne II] & 12.81 & 21.56 & 0.57 & 0.62 & 0.24 & 11.11 & 0.58 & 0.09 & 8.46 \\\relax
[Ne III] & 15.56 & 40.96 & 0.63  & 0.70 & 0.31 & 14.35 & 0.69 & 0.30 & 28.22 \\\relax
[S III] & 18.71 & 23.34 & 0.86 & 0.99 & 0.49 & 22.69 & 0.86 & --$^{\ddag}$ & -- \relax
\enddata

\tablecomments{ The FWHM of the MRS PSF (FWHM$_{\rm MRS}$) is taken from \cite{Arg2023}. We combine this with the measured spatial FWHM (${\rm FWHM}_{\rm spat}$) via Equation~\ref{eqn:fwhmspat} to calculate the corrected FWHM (${\rm FWHM}_{\rm spat,corr}$). We only report the lines that we were able to spatially resolve in at least one galaxy. See Section~\ref{sec:FWHM} for details.\\ $^{*}$ FWHM$_{\rm spat}$ measurement unavailable. \\$^\ddag$ Line is unresolved, FWHM$_{\rm spat}$ $<$ FWHM$_{\rm MRS}$.}

\end{deluxetable*}





\bibliography{reveallagn0}{}
\bibliographystyle{aasjournal}



\end{document}